\documentclass[twocolumn,pra,floatfix,amsmath,amsfonts]{revtex4}
\usepackage[latin9]{inputenc}
\usepackage{color}
\usepackage{amsmath}
\usepackage{amssymb}
\usepackage{graphicx}
\usepackage{esint}

\makeatletter
\@ifundefined{textcolor}{}
{%
 \definecolor{BLACK}{gray}{0}
 \definecolor{WHITE}{gray}{1}
 \definecolor{RED}{rgb}{1,0,0}
 \definecolor{GREEN}{rgb}{0,1,0}
 \definecolor{BLUE}{rgb}{0,0,1}
 \definecolor{CYAN}{cmyk}{1,0,0,0}
 \definecolor{MAGENTA}{cmyk}{0,1,0,0}
 \definecolor{YELLOW}{cmyk}{0,0,1,0}
}

\@ifundefined{textcolor}{}{%
 \definecolor{BLACK}{gray}{0}
 \definecolor{WHITE}{gray}{1}
 \definecolor{RED}{rgb}{1,0,0}
 \definecolor{GREEN}{rgb}{0,1,0}
 \definecolor{BLUE}{rgb}{0,0,1}
 \definecolor{CYAN}{cmyk}{1,0,0,0}
 \definecolor{MAGENTA}{cmyk}{0,1,0,0}
 \definecolor{YELLOW}{cmyk}{0,0,1,0}
}
\usepackage{graphics}\usepackage{epsfig}

\makeatother

\begin{document}

\title{Compacton matter waves in binary Bose gases under strong nonlinear
management 
}

\author{F.Kh. Abdullaev$^{1}$, M.S.A. Hadi$^{1}$, M. Salerno$^{2}$, and
B. Umarov$^{1}$ }

\affiliation{$^{1}$ Deparment of Physics, Kulliyyah of Science, International
Islamic University Malaysia, 25200 Kuantan, Pahang, Malaysia}

\affiliation{$^{2}$ Dipartimento di Fisica ``E.R. Caianiello'', CNISM and INFN
- Gruppo Collegato di Salerno, Universitá di Salerno, Via Giovanni
Paolo II, 84084 Fisciano (SA), Italy}

\date{\today}
\begin{abstract}
The existence of compacton matter waves in binary mixtures of quasi
one-dimensional Bose-Einstein condensates in deep optical lattices
and in the presence of nonlinearity management, is first demonstrated.
For this, we derive an averaged vector discrete nonlinear Schrödinger
equation (DNLSE) and show that compacton solutions of different types
can exist as stable excitations. Stability properties are studied
by linear analysis and by direct numerical integrations of the DNLSE
system and their dependence on the inter- and intra-species scattering
lengths, investigated. We show that under proper management conditions,
compactons can be very robust excitations that can emerge spontaneously
from generic initial conditions. A possible experimental setting for
compacton observation is also discussed.
\end{abstract}
\maketitle

\section{Introduction}

It has been recently demonstrated that Bose-Einstein condensates (BEC)
in deep optical lattices (OL) when exposed to strong and rapid periodic
time modulations of the scattering lengths can support matter wave
compactons, i.e. localized excitations with a compact support~\cite{AKS}.
Similarly to discrete breathers, compactons are intrinsically localized
stable excitations. In contrast with them, however, compactons have
no exponential tails, the lacking being due to the effective nonlinear
dispersion induced by the modulation, which permits the vanishing
of the tunneling just at the compacton edges. In particular, it was
shown (see~\cite{AKS} for details) that after averaging out the
fast time scale dynamics, the tunneling rate of the resulting averaged
system depends not only on modulation parameters but also on local
field density differences (atom numbers in the BEC case) between neighboring
sites. Thus, the periodic modulation in time of the scattering length
(nonlinear management) can be used to change the original dispersion
(e.g. the one in absence of management) into an effective nonlinear
dispersion essential for the compacton existence.

Field dependent tunneling suppression in the presence of strongly
and rapidly modulated interactions was demonstrated in~\cite{GMH}
for the case of a BEC trapped in a double well potential (compacton
formation being severely restricted in this case by the size of the
system) and in~\cite{AKS} for a one-dimensional BEC array modeled
by the discrete nonlinear Schrödinger equation (DNLSE). The phenomenon
of compacton formation, however, is of general validity and, as we
demonstrate in this paper, can occur also in more complicated BEC
systems.

From an experimental point of view, periodic time modulations of the
scattering length can be achieved by the Feshbach resonance (FR) technique~\cite{Inouye},
e.g. by varying the external magnetic field near a resonant value.
Besides compacton excitations, periodic modulations of the scattering
lengths can be used to create density dependent gauge fields~\cite{Greschner},
this being presently a field of rapidly growing interest, connected
with interesting physical phenomena, including pair superfluidity,
exactly-defect free Mott-insulator states etc.~\cite{Akos}. Properties
of the superfluid -Mott transition in a 2D square and a 3D cubic optical
lattice with periodic modulation of the atomic scattering length have
been investigated in~\cite{Wang}. Modulations of the interactions
were used also to design new correlated-hopping models for fermions
in optical lattices~\cite{LCJS} and, in combination with OL shaking,
for engineering unconventional Bose-Hubbard models~\cite{GSP}.

Spatial and temporal periodic changes of the scattering lengths have
been shown to be effective tools to change stability properties of
nonlinear excitations, leading to the existence of two dimensional
bright solitons in one and two component attractive condensates~\cite{SU01,ACMK,MPG,SKP,AG}.
Moreover, they were used to induce long lived Bloch oscillations~\cite{Salerno08,Gaul,Diaz},
dynamical localization~\cite{din-loc},  Rabi oscillations~\cite{rabi}
of BEC gap solitons in optical lattices, Faraday waves~\cite{LVS,AOS}
etc. Nonlinear management techniques were considered also in nonlinear
optics to stabilize 2D and 3D solitons and to reduce collapse in optically
layered media with self-focusing interaction~\cite{SU01}, to improve
communication capacities via soliton dispersion management in optical
fibers~\cite{Gabitov}, to create linear superpositions of gap solitons
in periodic Kerr media~\cite{sup-s}, etc.

All these studies refer to the single component case, e.g. BEC made
by a a single atomic species. An interesting question to ask, however,
is wether compacton excitations could exist also in multi-component
systems of interest both for BEC and nonlinear optics. The aim of
the present paper is just to provide an answer to this question.

In this respect we introduce an averaged vector DNLSE with effective
nonlinear discrete dispersion terms which describes the dynamics of
a BEC array in the presence of strong inter-species and intra-species
scattering length modulations. The existence of different compacton
states in the form of bright-bright (B-B), bright-dark (B-D), and
dark-dark (D-D) pairs, is first demonstrated and stability properties
of these states investigated both by means of a linear stability analysis
and by direct numerical integrations of the model equations. As a
result we find that while single site and two sites out of phase B-B
compactons are stable in the whole parameter range, for the other
modes there exists thresholds in the tunneling constant rate below
which they cannot exist as stable excitations. The dependence of the
compacton stability on inter-species interaction is also investigated.
For D-D compactons we find that when the inter-species scattering
length is detuned to zero (e.g. in the uncoupled single component
limit) the stability becomes narrower compared to that of the pure
two components case. The predictions of our analysis are shown to
be in good agreement with the results obtained direct numerical simulations
of the model. The emergence of compact excitations from generic initial
excitations and possible experimental settings for their observation
are also discussed.

The paper is organized as follows. In Sec. II we introduce the model
equations and discuss the averaged equations. In Section III we derive
the conditions for existence of two component of B-B, B-D and D-D
compactons and discuss stability properties. In section IV we consider
the case of D-D compactons in the uncoupled limit corresponding to
the single component dark compacton case. In section V the emergence
of compact excitations from generic initial excitations is investigated
and possible experimental parameter design provided. Finally, in Sec.
VI the results of the paper are briefly summarized.

\section{Model equations and averaging}

Two-component atomic BEC in a deep OL can be described in the tight
binding approximation by the following vector DNLSE~\cite{Shrestha}
\begin{eqnarray}
 & i\dot{u}_{n}=-\kappa_{1}(u_{n+1}+u_{n-1})-(\gamma_{1}|u_{n}|^{2}+\gamma_{12}|v_{n}|^{2})u_{n},\nonumber \\
\label{vdnlse}\\
 & i\dot{v}_{n}=-\kappa_{2}(v_{n+1}+v_{n-1})-(\gamma_{12}|u_{n}|^{2}+\gamma_{2}|v_{n}|^{2})v_{n},\nonumber
\end{eqnarray}
where the overdot stands for time derivative, the coefficients $\kappa_{i},\; i=1,2,$
are related to the tunneling rates of atoms between neighboring wells
of the optical lattice and $\gamma_{12}$, $\gamma_{i},\; i=1,2$
are nonlinear coefficients related to the inter-species ($a_{12}$)
and intra-species ($a_{ii},\; i=1,2$) scattering lengths, respectively.
Eq. (\ref{vdnlse}) arises also in nonlinear optics where they model
the propagation of the electric field in array of optical waveguides
with variable Kerr nonlinearity. In this context the role of time
is played by the longitudinal propagation distance along the optical
fiber and the nonlinear coefficients $\gamma_{i},\gamma_{12}$ correspond to
self- and cross-phase modulations of the electric field components,
respectively~\cite{Kobyakov,Assanto}. Notice that the above two
component DNLSE has the Hamiltonian form $\dot{\chi}_{n}=\delta H/\delta\chi_{n}^{*}$
with $\chi_{n}=u_{n},v_{n}$ and the Hamiltonian $H$ given by
\begin{equation}
\begin{split}H=- & \sum_{n}\Big[(\kappa_{1}u_{n+1}u_{n}^{\ast}+\kappa_{2}v_{n+1}v_{n}^{\ast}+c.c.)\,+\\
 & \frac{1}{2}\left(\gamma_{1}|u_{n}|^{4}+\gamma_{2}|v_{n}|^{4}\right)+\gamma_{12}|u_{n}|^{2}|v_{n}|^{2}\Big].\label{HOri}
\end{split}
\end{equation}
Here the $*$ stands for the complex conjugation and $c.c.$ denotes
the complex conjugate of the expression in the parenthesis. Also notice
that the number of atoms $N_{i}=\sum_{n}|\chi_{n}|^{2},\;\;\chi_{n}=u_{n},v_{n}$
are conserved for each component.

Taking into account that Eq. (\ref{vdnlse}) arises in the tight binding
approximation~\cite{TS,ABKS}, the nonlinear coefficients $\gamma_{i},\gamma_{12}$
can be expressed in terms of the overlap integrals of Wannier $W_{n}^{(i)}(x)$
functions of the corresponding continuous periodic Gross-Pitaevskii
mean field model~\cite{AKKS} as:
\begin{eqnarray*}
 &  & \gamma_{12}=\frac{2\pi\hbar\sqrt{N_{1}N_{2}}a_{12}}{m}\int dx|W_{n}^{(1)}|^{2}|W_{n}^{(2)}|^{2},\\
 &  & \gamma_{i}=\frac{4\pi\hbar N_{i}a_{ii}}{m_{i}}\int dx|W_{n}^{i}|^{4},\;\; i=1,2,
\end{eqnarray*}
with $m_{i},\; i=1,2$ denoting the atomic masses and $m=m_{1}m_{2}/(m_{1}+m_{2})$
the reduced mass (functions $W_{n}$ are normalized on the whole line:
$\int|W_{n}^{i}(x)|^{2}dx=1$). The modulations of the nonlinear coefficients
$\gamma_{i},\gamma_{12}$ is assumed to be of the form
\begin{equation}
\gamma_{i}(t)=\gamma_{i}^{(0)}+\frac{1}{\epsilon}\,\gamma_{i}^{(1)}\left(\frac{t}{\epsilon}\right),
\end{equation}
with $\gamma^{(0)}$ a constant, $\gamma^{(1)}(t)$ a rapidly varying
periodic function of time and $\epsilon$ a small parameter that controls
the strength and the frequency of the modulation (strong nonlinearity
management corresponds to $\epsilon\ll1$). For $\gamma_{i}^{(1)}(t)$
we consider periodic modulations of the type
\begin{equation}
\gamma_{i}^{(1)}(t)=\frac{\gamma_{i}^{(1)}}{\epsilon}\cos(\Omega\frac{t}{\epsilon}),\;\;\;\ \Omega,\;\gamma_{i}^{(1)}\sim O(1)
\end{equation}
of period $T=2\pi/\Omega$ in the fast time variable $\tau=t/\epsilon$.
Although in general one could consider independent modulations of
the nonlinear coefficients, in practical contexts one usually deals
with the simpler settings
\begin{itemize}
\item {i)} $\gamma_{i}=\gamma_{i}\left(t/\epsilon\right),\;\;\;\;\;\;\;\gamma_{12}=\text{const}$,\;\;
i=1,2 \\

\item {ii)} $\gamma_{12}=\gamma_{12}\left(t/\epsilon\right),\;\;\;\gamma_{i}=\text{const}$,\;\;
i=1,2.
\end{itemize}
In the BEC context, case \textit{i} corresponds to a modulation in
time of the intra-species scattering lengths, keeping inter-species
scattering length constant, while in case \textit{ii} it is done just
the opposite.  Although these settings are both feasible
for BEC mixtures, case {ii)} is not easy to implement in nonlinear
optics because in this context cross-phase modulations are usually
difficult to control. Moreover, the averaged equations obtained in
the case {ii)} are mathematically more involved due to the presence
of a  complicated nonlinear dispersion (see below). In the following
we concentrate mainly on case {i)} and discuss case {ii)} only
briefly at the end of the section.

Let us consider then a fixed inter-species scattering length ($\gamma_{12}=\text{const}$)
and assume the intra-species scattering lengths modulated as follows:
\begin{equation}
\gamma_{1}=\gamma_{1}^{(0)}+\gamma_{1}^{(1)}\left(t/\epsilon\right),\gamma_{2}=\gamma_{2}^{(0)}+\gamma_{2}^{(1)}\left(t/\epsilon\right).
\end{equation}
To find effective nonlinear evolution equations, we use averaging
method to eliminate the fast time, $\tau=t/\epsilon$, dependence.
In this respect it is convenient to perform the following transformation
\begin{equation}
u_{n}=U_{n}e^{i\Gamma_{1}|U_{n}|^{2}},\; v_{n}=V_{n}e^{i\Gamma_{2}|V_{n}|^{2}},\label{eq2}
\end{equation}
where $\Gamma_{i}$ are the antiderivatives of $\gamma_{i}^{(1)}(t)$,
$$
\Gamma_i(t)=\int_0^t \gamma_i^{(1)}(t^{\prime})dt^{\prime} - \frac{1}{T}\int_0^T\int_0^t\gamma_i^{(1)}(t^{\prime}) dt^{\prime}dt,
$$
where $T$ is the period.
By substituting Eq.(\ref{eq2}) into Eqs.(\ref{vdnlse}) we obtain:
\begin{equation}
\begin{split}i\dot{U}_{n}= & \, i\kappa_{1}\Gamma_{1}(\tau)U_{n}[U_{n}^{\ast}X_{1}-U_{n}X_{1}^{*}]-\\
 & \,\kappa_{1}X_{1}-(\gamma_{1}^{(0)}|U_{n}|^{2}+\gamma_{12}|V_{n}|^{2})U_{n}
\end{split}
\end{equation}
\begin{equation}
\begin{split}i\dot{V}_{n}= & \, i\kappa_{2}\Gamma_{2}(\tau)V_{n}[V_{n}^{\ast}X_{2}-V_{n}X_{2}^{*}]-\\
 & \,\kappa_{2}X_{2}-(\gamma_{2}^{(0)}|V_{n}|^{2}+\gamma_{12}|U_{n}|^{2})V_{n}\label{eq.tr}
\end{split}
\end{equation}
where $X_{1}=U_{n+1}e^{i\Gamma_1\theta_{1}^{+}}+U_{n-1}e^{i\Gamma_1\theta_{1}^{-}}$,
$X_{2}=V_{n+1}e^{i\Gamma_2\theta_{2}^{+}}+V_{n-1}e^{i\Gamma_2\theta_{2}^{-}}$ and
\begin{equation}
\theta_{1}^{\pm}=|U_{n\pm1}|^{2}-|U_{n}|^{2}, \theta_{2}^{\pm}=|V_{n\pm1}|^{2}-|V_{n}|^{2}.
\end{equation}
The average over the rapid modulation can be done with the help of
the relations
\begin{equation}
\begin{split}<e^{\pm i\Gamma_{i}\theta^{\pm}}> & =J_{0}(\alpha_{i}\theta^{\pm})\\
<\Gamma_{i}e^{\pm i\Gamma_i\theta^{\pm}}> & =\pm i\,\alpha_{i}J_{1}(\alpha_{i}\theta^{\pm}),
\end{split}
\end{equation}
where $J_{0}$, $J_{1}$ are Bessel functions~\cite{abramowitz}
of first kind of zero-th and first order, respectively, and with $\alpha_{i}$
given by
\begin{equation}
\alpha_{i}=\gamma_{i}/\Omega,\ i=1,2. \label{alfai}
\end{equation}

Here the angular bracket denotes average with respect to the fast
time variable, e.g. $<F>\equiv(1/T)\int_{0}^{T}Fd\tau$. The system
of averaged equations is then obtained as:
\begin{equation}
\begin{split}i\dot{U}_{n}=- & \alpha_{1}\kappa_{1}U_{n}\Big[J_{1}(\alpha_{1}\theta_{1}^{+})\,\left(U_{n}^{\ast}U_{n+1}+U_{n}U_{n+1}^{\ast}\right)\\
 & \qquad\,+J_{1}(\alpha_{1}\theta_{1}^{-})\,\left(U_{n}^{\ast}U_{n-1}+U_{n}U_{n-1}^{\ast}\right)\Big]\\
- & \,\kappa_{1}\left[U_{n+1}J_{0}(\alpha_{1}\theta_{1}^{+})+U_{n-1}J_{0}(\alpha_{1}\theta_{1}^{-})\right]\\
- & \left[\gamma_{1}^{(0)}|U_{n}|^{2}+\gamma_{12}|V_{n}|^{2}\right]U_{n},
\end{split}
\label{eq.av1}
\end{equation}
\begin{equation}
\begin{split}i\dot{V}_{n}=- & \alpha_{2}\kappa_{2}V_{n}\Big[J_{1}(\alpha_{2}\theta_{2}^{+})\,\left(V_{n}^{\ast}V_{n+1}+V_{n}V_{n+1}^{\ast}\right)\\
 & \qquad\,+J_{1}(\alpha_{2}\theta_{2}^{-})\,\left(V_{n}^{\ast}V_{n-1}+V_{n}V_{n-1}^{\ast}\right)\Big]\\
- & \,\kappa_{2}\left[V_{n+1}J_{0}(\alpha_{2}\theta_{2}^{+})+V_{n-1}J_{0}(\alpha_{2}\theta_{2}^{-})\right]\\
- & \left[\gamma_{2}^{(0)}|V_{n}|^{2}+\gamma_{12}|U_{n}|^{2}\right]V_{n}.
\end{split}
\label{eq.av2}
\end{equation}

Note that Eqs. (\ref{eq.av1}, \ref{eq.av2}) have Hamiltonian form
with averaged Hamiltonian, $H_{av}$, given by: \
\begin{widetext}
\begin{equation}
H_{av}=-\sum_{n}\Big[\kappa_{1}J_{0}(\alpha_{1}\theta_{1}^{+})\,[U_{n+1}U_{n}^{\ast}+c.c.]+\kappa_{2}J_{0}(\alpha_{2}\theta_{2}^{+})[V_{n+1}V_{n}^{\ast}+c.c.]+\frac{1}{2}\left(\gamma_{1}^{(0)}|U_{n}|^{4}+\gamma_{2}^{(0)}|V_{n}|^{4}\right)+\gamma_{12}|U_{n}|^{2}|V_{n}|^{2}\Big]\label{Hav}
\end{equation}
\end{widetext} By comparing Eq. (\ref{Hav}) with the corresponding
unperturbed Hamiltonian Eq. (\ref{HOri}), one can see that the effect
of the scattering lengths modulation simply reflects in the following nonlinear
rescaling of the tunneling constants:
\begin{equation}
\kappa_{i}\rightarrow\kappa_{i}J_{0}(\alpha_{i}\theta_{i}^{+}),\ i=1,2.\label{rescaling}
\end{equation}

From this equation and from Eq. (\ref{alfai}) it is clear that the
inter-species interaction $\gamma_{12}$ play no role in determining
the lattice sites where the zero tunneling condition occur (compacton
boundaries) when the nonlinear management is made with respect to
the intra-species interactions (see below for explicit examples).
Thus, in analogy with the single component case considered in~\cite{AKS}
the tunneling constants depend on the atom difference between neighboring
sites~\cite{GMH,AKS}. This introduces an effective nonlinear dispersion
in the system which leads to the existence of two-component compactons.

In closing this section it is interesting to discuss
the changes in the above derivation implied by a nonlinear management
of the inter-species scattering length
\begin{equation}
\gamma_{12}=\gamma_{12}^{(0)}+\gamma_{12}^{(1)}(t),
\end{equation}
with constant (not necessarily equal) intra-species parameters $\gamma_{1}$,
$\gamma_{2}$. In this case, to remove the explicit time dependence
from Eq. (\ref{vdnlse}), the transformation in (\ref{eq2}) must
be replaced by
\begin{equation}
u_{n}=U_{n}e^{i\Gamma(t)|V_{n}|^{2}},\; v_{n}=V_{n}e^{i\Gamma(t)|U_{n}|^{2}},\label{eq2bis}
\end{equation}
with $\Gamma(t)$ the antiderivative of $\gamma_{12}^{(1)}(t)$. Following the same approach as before, it is not difficult to show
that one arrives at the same averaged Hamiltonian as in Eq.~(\ref{Hav})
but with the interchanges $\kappa_{1}\leftrightarrow\kappa_{2}$,
$U\leftrightarrow V$, operated in the tunneling terms and the obvious
replacements $\gamma_{12}\rightarrow\gamma_{12}^{(0)}$, $\gamma_{i}^{(0)}\rightarrow\gamma_{i}$.
This implies a different rescaling of the tunneling constants
\begin{equation}
\kappa_{i}\rightarrow\kappa_{i}J_{0}(\alpha\theta_{3-i}^{+}),\;\;\;\;\;\;\; i=1,2,\label{rescaling}
\end{equation}
and leads to  complicated nonlinear dispersion terms in the averaged
equations. In view of this complexity, in the following we restrict
only to compactons induced by intra-species management. A detailed
study of the inter-species management requires more investigations
and will be discussed elsewhere \cite{AHSU}.

\section{Existence and stability of compactons in binary BEC mixtures}

Exact compacton solutions of the averaged system can be searched
as stationary states of the form:
\begin{equation}
U_{n}=A_{n}e^{-i\mu_{u}t},\ V_{n}=B_{n}e^{-i\mu_{v}t},
\end{equation}
with $\mu_{u},\mu_{v}$ chemical potentials of the two atomic species.
Substituting these expressions into Eq. (\ref{eq.av1},\ref{eq.av2})
one gets the following stationary equations:

\begin{eqnarray}
\mu_{u}A_{n}+(\gamma_{1}A_{n}^{3}+\gamma_{12}B_{n}^{2}A_{n})+\kappa_{1}[A_{n+1}J_{0}(\alpha_{1}\theta_{1}^{+})+\nonumber \\
A_{n-1}J_{0}(\alpha_{1}\theta_{1}^{-})]+2\alpha_{1}\kappa_{1}A_{n}^{2}[A_{n+1}J_{1}(\alpha_{1}\theta_{1}^{+})+\label{mubg1}\\
A_{n-1}J_{1}(\alpha_{1}\theta_{1}^{-})]=0,\nonumber \\
\nonumber \\
\mu_{v}B_{n}+(\gamma_{2}B_{n}^{3}+\gamma_{12}A_{n}^{2}B_{n})+\kappa_{2}[B_{n+1}J_{0}(\alpha_{2}\theta_{2}^{+})+\nonumber \\
B_{n-1}J_{0}(\alpha_{2}\theta_{2}^{-})]+2\alpha_{2}\kappa_{2}B_{n}^{2}[B_{n+1}J_{1}(\alpha_{2}\theta_{2}^{+})+\label{mubg2}\\
B_{n-1}J_{1}(\alpha_{2}\theta_{2}^{-})]=0,\nonumber
\end{eqnarray}
to be solved for the chemical potentials and amplitudes $A_{n},B_{n}$
of the compacton modes. The compact nature of the solution ($A_{i},B_{i}=0$
outside a finite (small) range of sites), allows to truncate the above
infinite system into a finite number of relations between the above
variables, which can be solved exactly. In the following subsections
this is shown explicitly for the different compacton types.
\begin{figure}
\centerline{\includegraphics[scale=0.5]{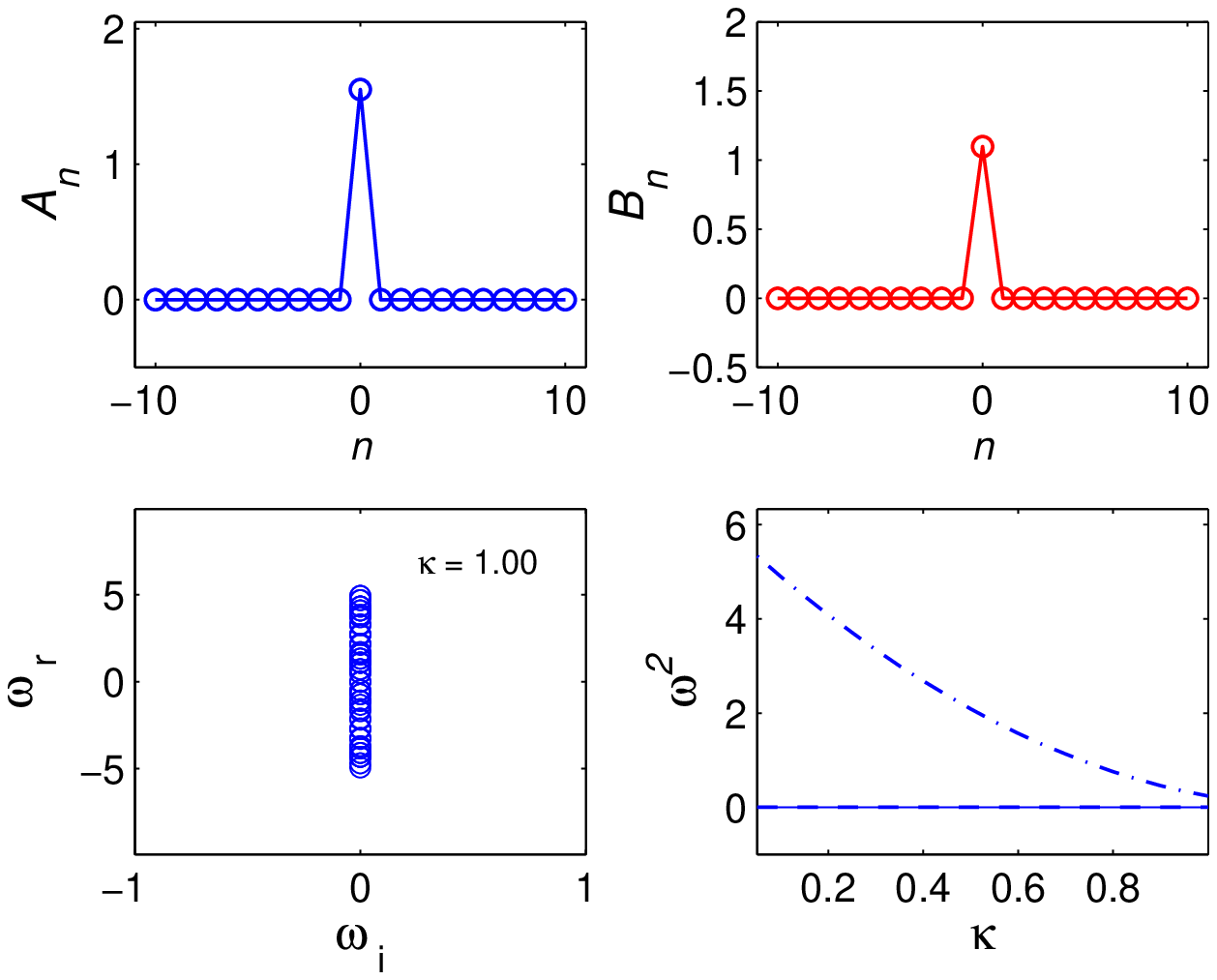}} \centerline{\includegraphics[scale=0.5]{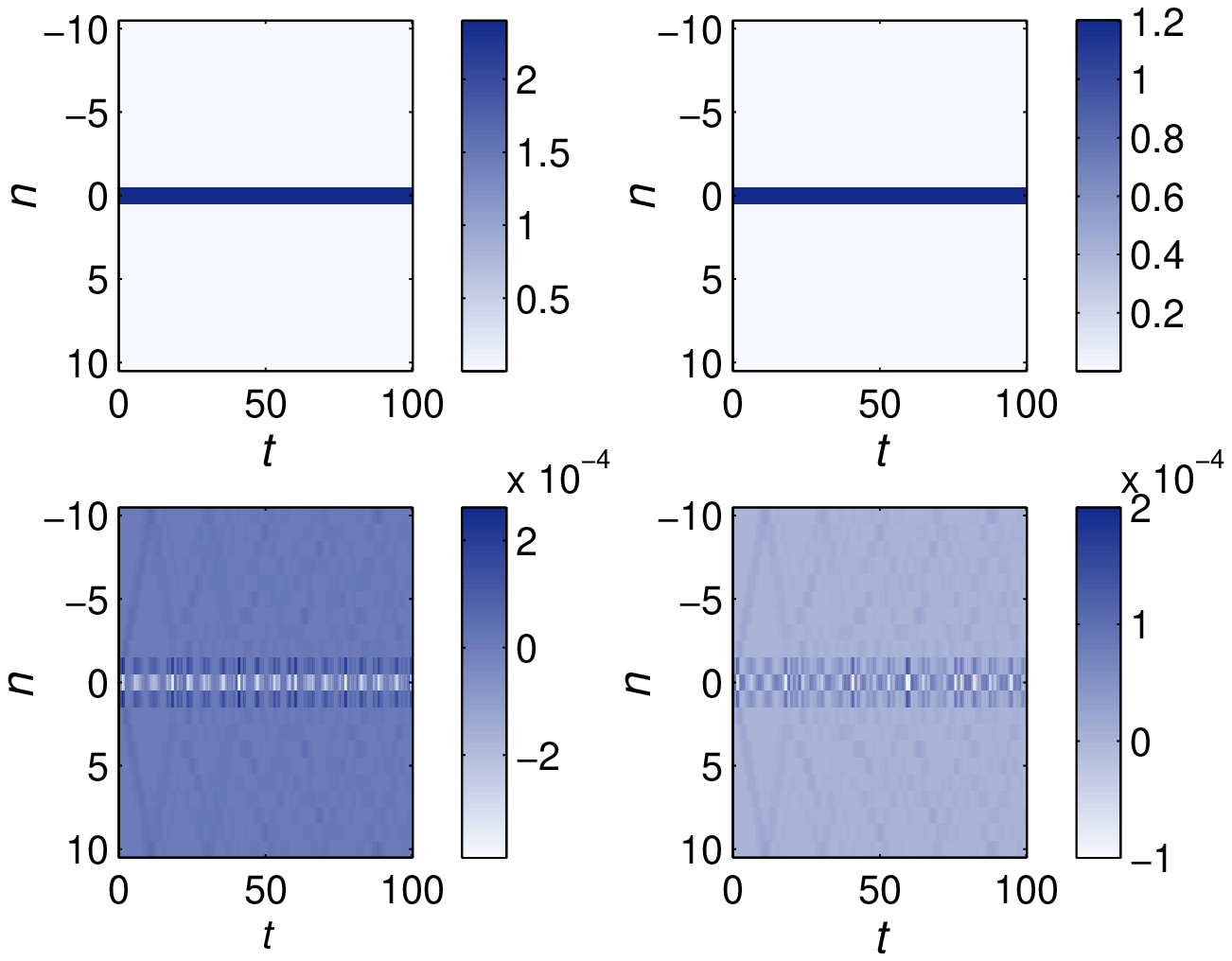}}
\protect\caption{(Color online) First row panels. Exact single site B-B compacton of
Eqs. (\ref{eq.av1}-\ref{eq.av2}) for $\gamma_{1}=\gamma_{2}=1$,
$\gamma_{12}=0.5$, $\kappa_{1}=\kappa_{2}\equiv\kappa=1$, $\alpha_{1}=1$,
and $\alpha_{2}=2$. Second row panels. Real and imaginary parts of
the eigenfrequency spectrum (left panel) corresponding to the B-B
compacton depicted in top row panels. Right panel shows the lowest
(solid line) third lowest (dash line) and fifth lowest (dash-dot line)
$\omega^{2}$ values as a function of $\kappa$. Third row panels.
Space-time evolution of $|A_{n}|^{2}$ (left panel) and $|B_{n}|^{2}$
(right panel) as obtained from direct numerical integration of Eq.
(\ref{vdnlse}) with $\kappa_{i}=0.5,\;\gamma_{i}=1+\frac{\alpha_{i}}{\epsilon}\cos\left(t/\epsilon\right),i=1,2$,
$\epsilon=0.01$, taking as initial condition the corresponding exact
single site B-B compacton of the averaged system (other parameters
are fixed as for top row panels). Bottom row panel. Deviation of the
dynamics depicted in third row panels from the corresponding one obtained
from Eqs. (\ref{eq.av1}-\ref{eq.av2}).}

\label{brightbrightsingle}
\end{figure}

\subsection{Bright-Bright compactons}

To search  compacton solutions of the B-B type  we need to look
for the last sites of vanishing amplitude, say $n_{0}\pm1$, where
the vanishing of the tunneling rate is realized. For a single site
B-B compacton we assume $A_{n_{0}}=a,\, B_{n_{0}}=b,\; A_{n_{0}\pm j}=0,\, B_{n_{0}\pm j}=0$
for all $j\ge1$. Substituting this ansatz in Eqs.(\ref{mubg1},\ref{mubg2})
we obtain the corresponding condition for the compacton existence
as
\begin{eqnarray}
J_{0}(\alpha_{1}a^{2})=0,\ a^{2}=\xi_{0}/\alpha_{1},\nonumber \\
J_{0}(\alpha_{2}b^{2})=0,\ b^{2}=\xi_{0}/\alpha_{2}\label{aconst}
\end{eqnarray}
where $\xi_{0}$ is a zero of the Bessel function $J_{0}$ (in all
numerical calculations below we take the first zero of $J_{0}$: $\xi_{0}=2.4048$).
This condition together with
\begin{equation}
\mu_{u}=-\gamma_{1}a^{2}-\gamma_{12}b^{2},\ \mu_{v}=-\gamma_{2}b^{2}-\gamma_{12}a^{2}
\end{equation}
gives us the single site B-B compacton pair.

\begin{figure}
\centerline{\includegraphics[scale=0.5]{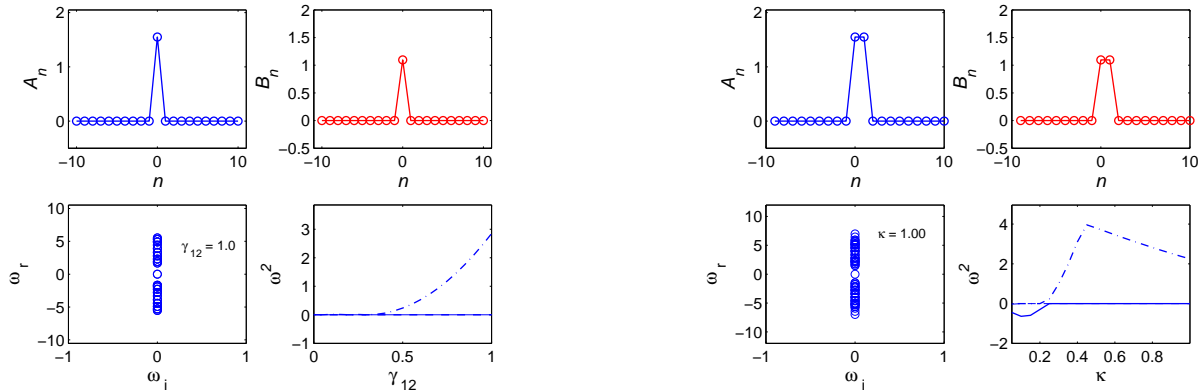}} \protect\caption{(Color online) Same as first two rows panels of Fig.~\ref{brightbrightsingle},
except for compacton in top panels computed for $\gamma_{12}=1$ and
stability properties in the bottom right panel computed as a function
of $\gamma_{12}$. Other parameters are fixed as in Fig.~\ref{brightbrightsingle}. }

\label{BBg12}
\end{figure}

Typical examples of single site B-B compactons are depicted in top
panels of Figs.~ \ref{brightbrightsingle},\ref{BBg12}. Stability
properties of the solution have been investigated by standard linear
analysis. Denoting by $\omega$ the eigenfrequencies of the linearized
averaged equations, associated to growing perturbations of the form
$e^{-i\omega t}$, we have that linear stability is granted if all
$\omega$ have zero imaginary parts. For the considered single site
B-B compacton, this condition is well satisfied, as one can see from
the left panel in the second row of Fig.~\ref{brightbrightsingle}.
Actually we find that single site B-B compactons are generically stable
for wide range of parameters. This can be seen from the second top
right panel of Fig.~\ref{brightbrightsingle} where the dependence
of the lowest squared eigenfrequencies, $\omega^{2}$, is reported
as a function of $\kappa$ (to reduce number of parameters we fix,
here and in the following , $\kappa_{1}=\kappa_{2}\equiv\kappa$)
. Extensive numerical linear stability studies show that $\omega^{2}$
is always non negative, meaning that the solution is linearly stable.

Nonlinear stability properties have been investigated by direct numerical
integrations of Eqs. (\ref{eq.av1}-\ref{eq.av2}) taking as initial
conditions exact compactons perturbed with a random uniformly distributed
noise field of amplitude $10^{-4}$. Results are found in full agreement
with the linear stability results discussed above. Excellent agreement
is also obtained from direct numerical integrations of the original
vector DNLSE (\ref{vdnlse}), as one can see from third and fourth
row panels of Fig.~\ref{brightbrightsingle}. In particular notice
from the bottom row panels that the deviation of the original dynamics
from the exact averaged dynamics is very small (four orders of magnitude
small for $\epsilon=0.01$) and can be made even smaller by further
decreasing $\epsilon$.
\begin{figure}
\centerline{\includegraphics[scale=0.5]{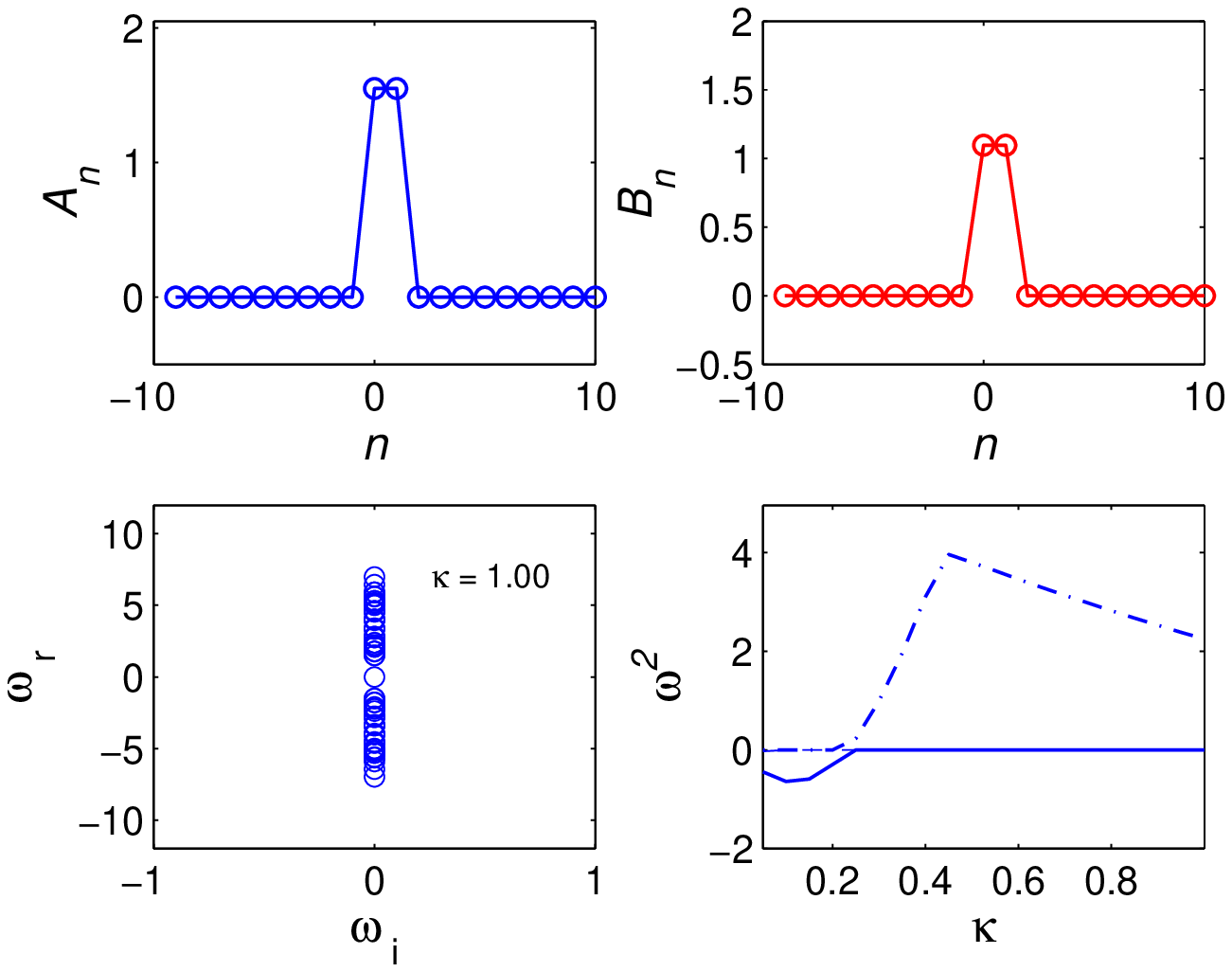}} \centerline{\includegraphics[scale=0.5]{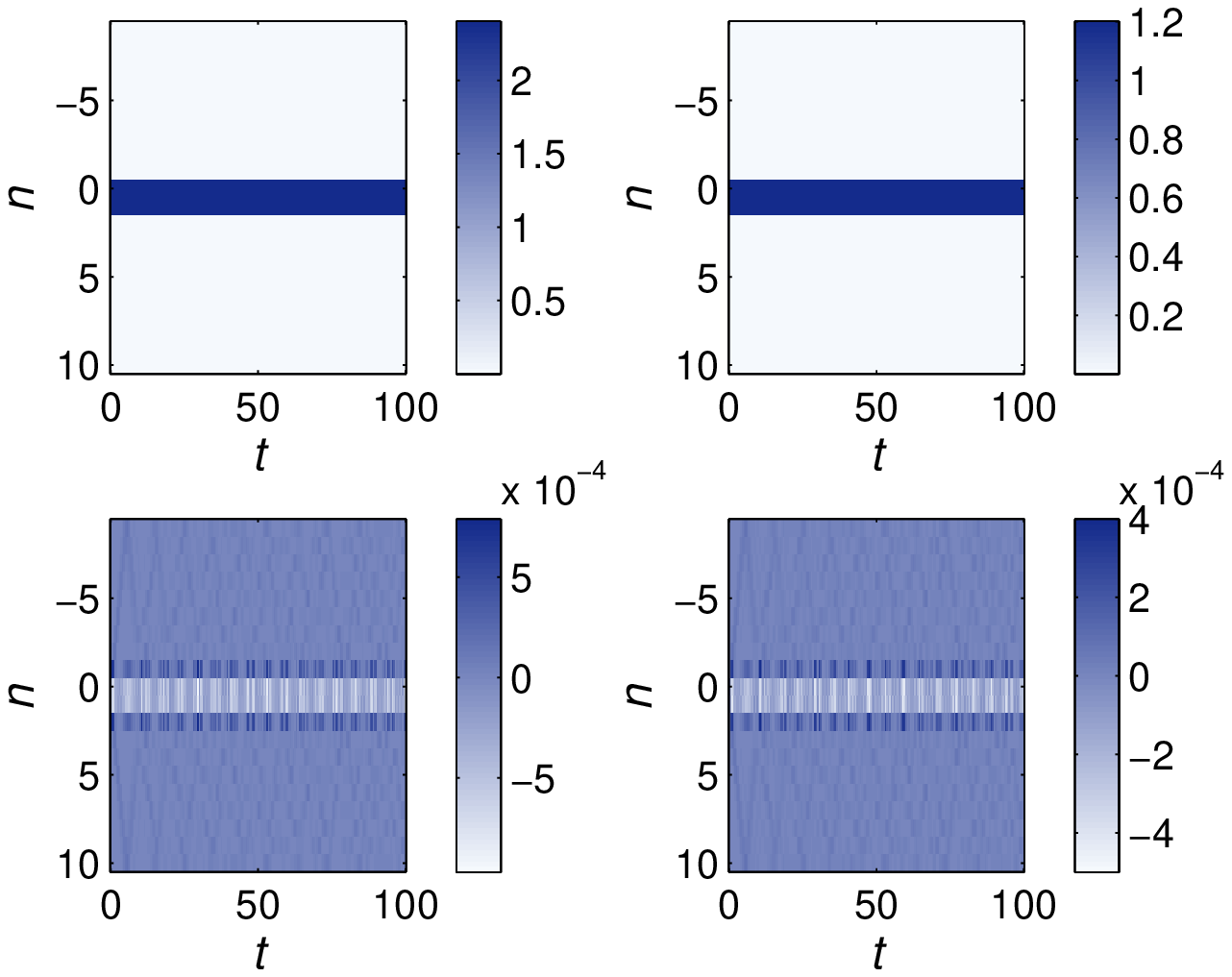}}
\protect\caption{(Color online) Same as in Fig.~\ref{brightbrightsingle} but for
a two-site in-phase B-B compacton. All parameters are fixed as in
Fig.~\ref{brightbrightsingle} }

\label{brightbrightwosite}
\end{figure}

In Fig.~\ref{BBg12} stability properties of single site B-B compactons
have been investigated as a function of the inter-species interaction,
$\gamma_{12}$, for a particular choice of the remaining parameters.
As one can see, the obtained behavior is very similar to the one obtained
for the $\kappa$ dependence in Fig.~\ref{brightbrightsingle}, this
further confirming the robustness of the one site B-B solution with
respect to wide parameters variations.

Two-site B-B compactons can also be found and, similarly to discrete
breathers, can be of two types: \textit{in-phase} and of \textit{out-of-phase}.
Two-site in-phase compactons follow from the ansatz: $A_{n_{0}}=a,\, A_{n_{0}-1}=0,\, A_{n_{0}+1}=a,\, B_{n_{0}}=b,\, B_{n_{0}-1}=0,\, B_{n+1}=b$.
By substituting into Eqs.(\ref{mubg1},\ref{mubg2}) one gets Eq.
(\ref{aconst}) as before with chemical potentials given by:
\begin{align*}
\mu_{u} & =-\gamma_{1}a^{2}-\gamma_{12}b^{2}-\kappa_{1},\\
\mu_{v} & =-\gamma_{2}b^{2}-\gamma_{12}a^{2}-\kappa_{2.}
\end{align*}
First two row panels of Fig.~\ref{brightbrightwosite} show a typical
two-site in-phase B-B stationary compacton. Notice from the second
row panels that the linear eigenfrequencies are all positive in the
interval $\left(0.25,1\right)$ this implying a wide stability rate also
in this case. Direct
numerical integrations of the averaged equation is shown in last two
rows of Fig.~\ref{brightbrightwosite}.

Out-phase B-B compactons can be obtained from the ansatz $A_{n_{0}}=a,\, A_{n_{0}-1}=0,\, A_{n_{0}+1}=-a,\, B_{n_{0}}=b,\, B_{n_{0}-1}=0,\, B_{n+1}=-b$.
In this case the chemical potentials are given by
\begin{align*}
\mu_{u} & =-\gamma_{1}a^{2}-\gamma_{12}b^{2}+\kappa_{1},\\
\mu_{v} & =-\gamma_{2}b^{2}-\gamma_{12}a^{2}+\kappa_{2},
\end{align*}
with $a,b,$ fixed as in Eq. (\ref{aconst}).
\begin{figure}
\centerline{\includegraphics[scale=0.5]{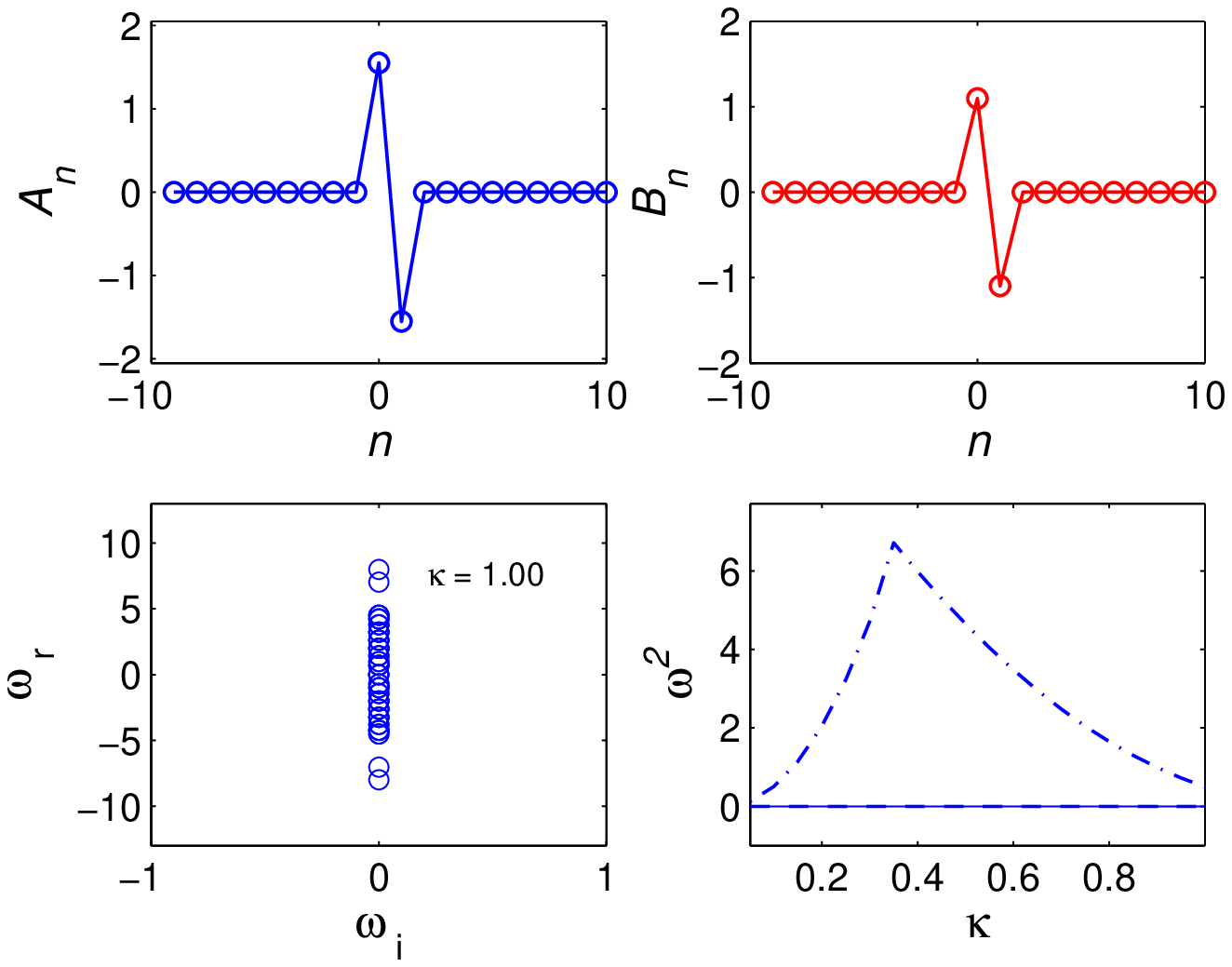}} \centerline{\includegraphics[scale=0.5]{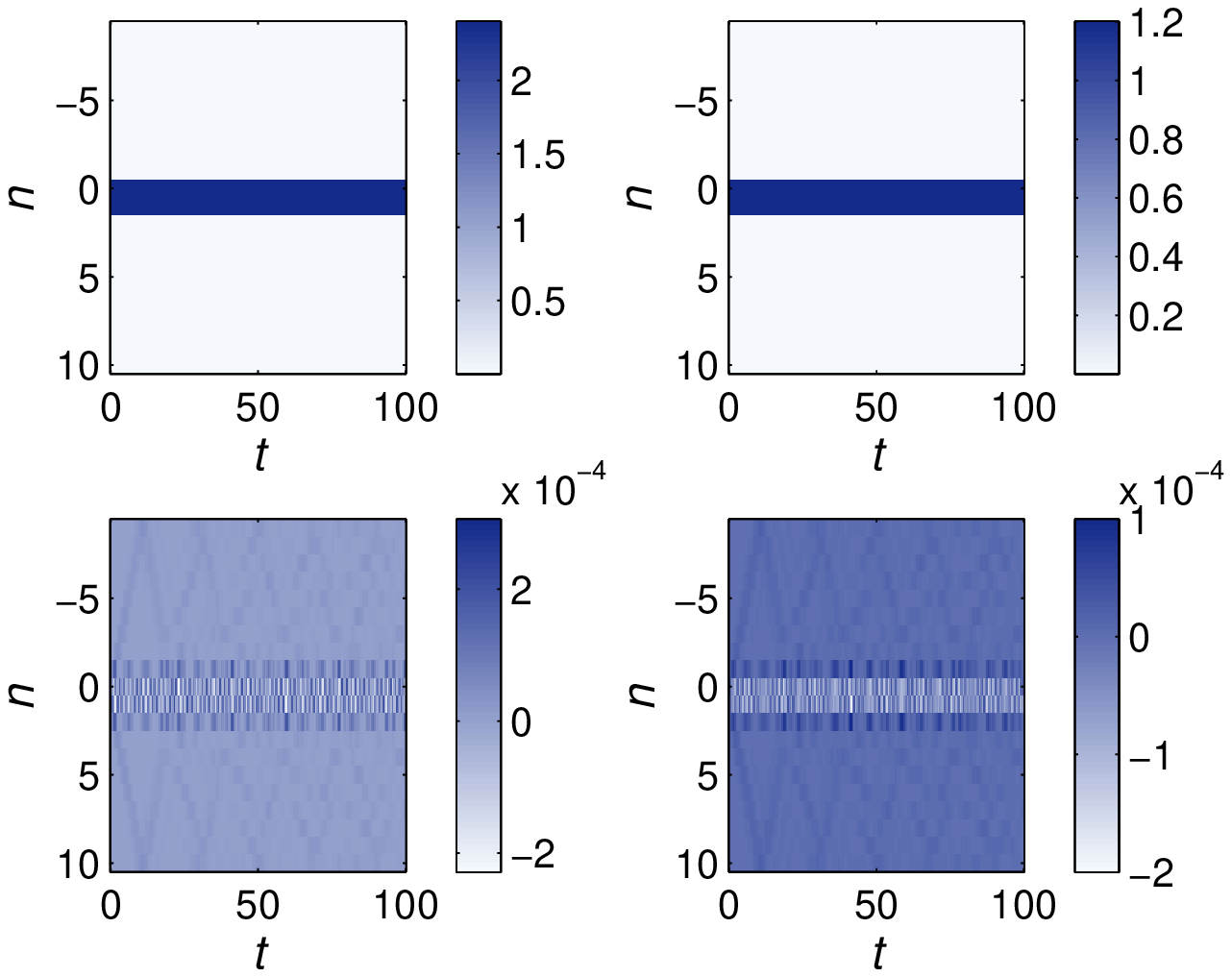}}
\protect\caption{(Color online) Same as in Fig.~\ref{brightbrightsingle} but for
a two-site out-of-phase B-B compacton. All parameters are fixed as
in Fig.~\ref{brightbrightsingle}. }

\label{bbtwositeoutkp}
\end{figure}
Typical two-site out-of-phase compacton and related linear stability
analysis and time evolutions, are reported in Fig.~\ref{bbtwositeoutkp}.

Quite remarkably it is also possible to find exact three-site B-B
compactons. In this case assuming $A_{n_{0}}=a_{1},\, A_{n_{0}\pm1}=a_{2},\, A_{n_{0}\pm2}=0,\, B_{n_{0}}=b_{1},\, B_{n_{0}\pm1}=b_{2},\, B_{n\pm2}=0$,
we obtain chemical potentials as
\begin{align*}
\mu_{u}= & -\gamma_{1}a_{2}^{2}-\gamma_{12}b_{2}^{2}-\kappa_{1}a_{1}J_{0}\left(\xi_{u}\right)/a_{2}\\
 & -2\alpha_{1}\kappa_{1}a_{1}a_{2}J_{1}\left(\xi_{u}\right)\\
\mu_{v}= & -\gamma_{2}b_{2}^{2}-\gamma_{12}a_{2}^{2}-\kappa_{2}b_{1}J_{0}\left(\xi_{v}\right)/b_{2}\\
 & -2\alpha_{2}\kappa_{2}b_{1}b_{2}J_{1}\left(\xi_{v}\right)
\end{align*}
where $\xi_{u}=\alpha_{1}\left(a_{1}^{2}-a_{2}^{2}\right)$ and $\xi_{v}=\alpha_{1}\left(b_{1}^{2}-b_{2}^{2}\right)$.
\begin{figure}
\centerline{\includegraphics[scale=0.5]{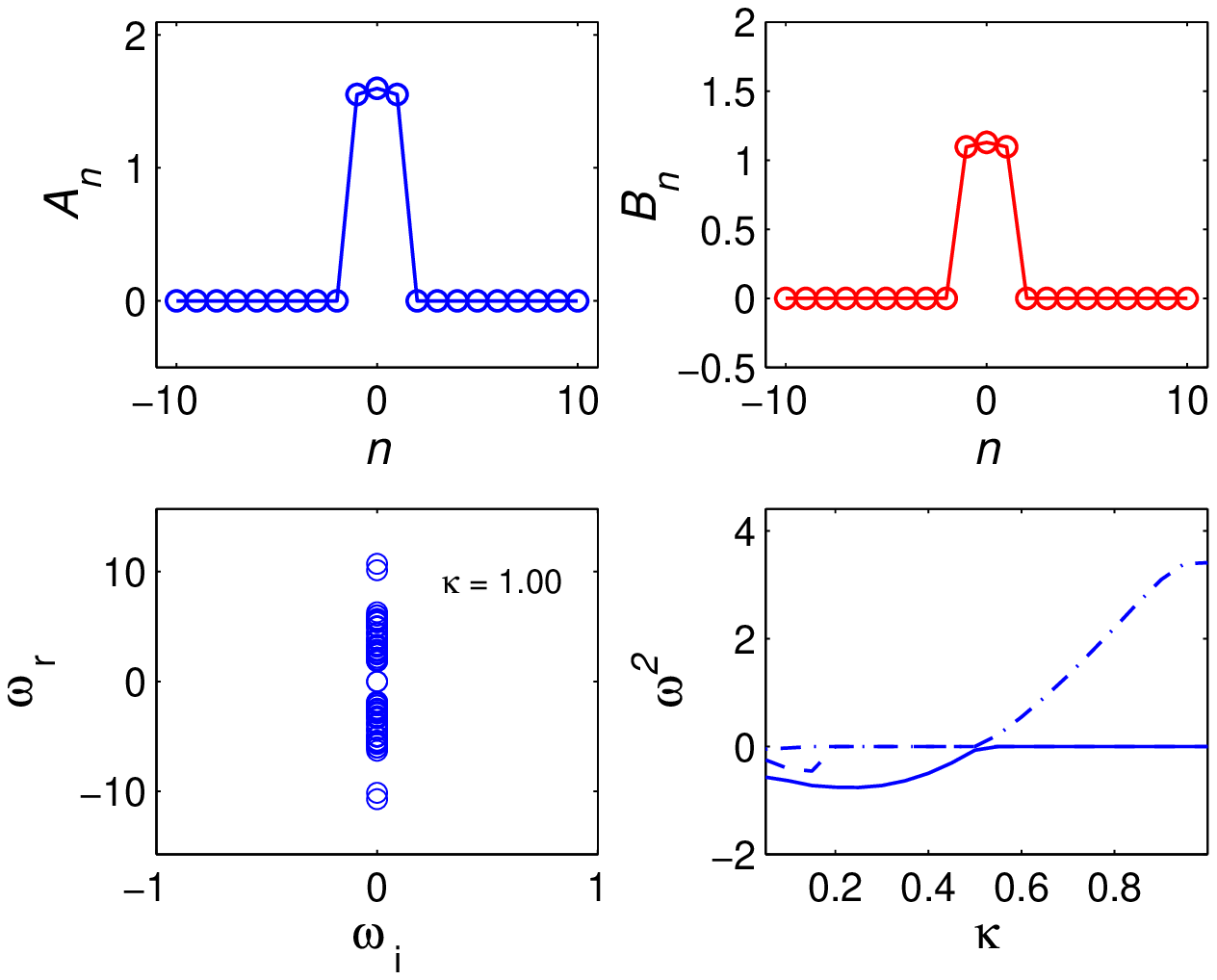}} \centerline{\includegraphics[scale=0.5]{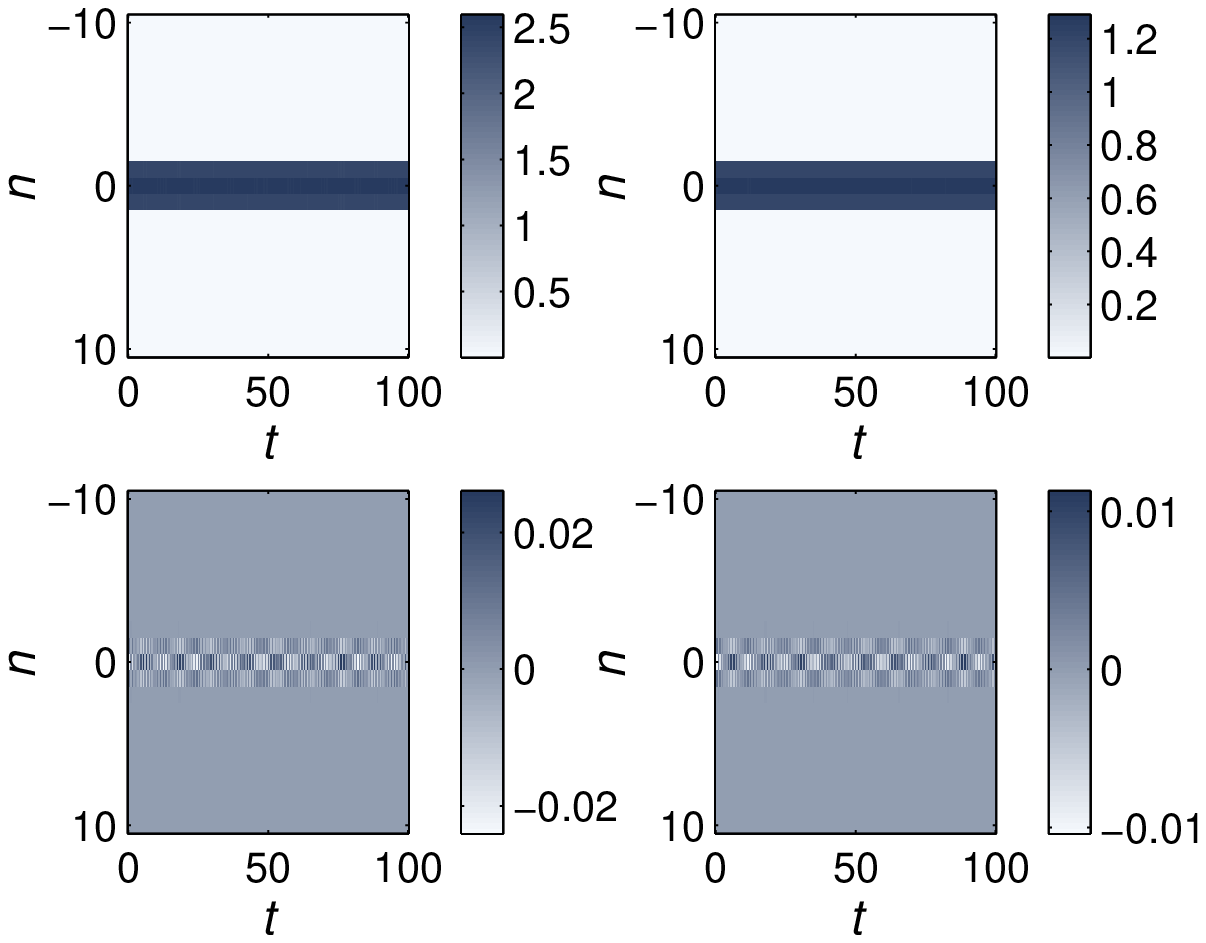}}
\protect\caption{(Color online) Same as in Fig.~\ref{brightbrightsingle} but for
a three-site B-B compacton. All parameters are fixed as in Fig.~\ref{brightbrightsingle}. }

\label{threebbkap}
\end{figure}

The constraint equations for the amplitudes are Eq.(\ref{aconst})
and
\begin{eqnarray*}
\gamma_{1}a_{1}a_{2}\left(a_{2}^{2}-a_{1}^{2}\right)+\gamma_{12}a_{1}a_{2}\left(b_{2}^{2}-b_{1}^{2}\right)\\
+\kappa_{1}\left(2a_{1}^{2}-a_{2}^{2}\right)J_{0}\left(\xi_{u}\right)+6\alpha_{1}\kappa_{1}a_{1}^{2}a_{2}^{2}J_{1}\left(\xi_{u}\right)=0,\\
\gamma_{2}b_{1}b_{2}\left(b_{2}^{2}-b_{1}^{2}\right)+\gamma_{12}b_{1}b_{2}\left(a_{2}^{2}-a_{1}^{2}\right)\\
+\kappa_{2}\left(2b_{1}^{2}-b_{2}^{2}\right)J_{0}\left(\xi_{v}\right)+6\alpha_{2}\kappa_{2}b_{1}^{2}b_{2}^{2}J_{1}\left(\xi_{v}\right)=0.
\end{eqnarray*}
Results are depicted in Fig.~\ref{threebbkap} for typical example.
Notice from second row right panel, that stability range in this case
is strongly reduced and stability is possible only for $\kappa>\approx0.5$.
Also notice from the bottom panels that the deviation
of the original dynamics relatively higher than the previous cases
of B-B. The amplitude profile differs mainly at the middle point but
this discrepancy, however, stays bounded in time and is quite small
if compared to the maximum amplitudes of the exact solution for each
component (for the considered case it never exceed the $2\%$ of the
exact amplitudes).
\begin{figure}
\centerline{\includegraphics[scale=0.5]{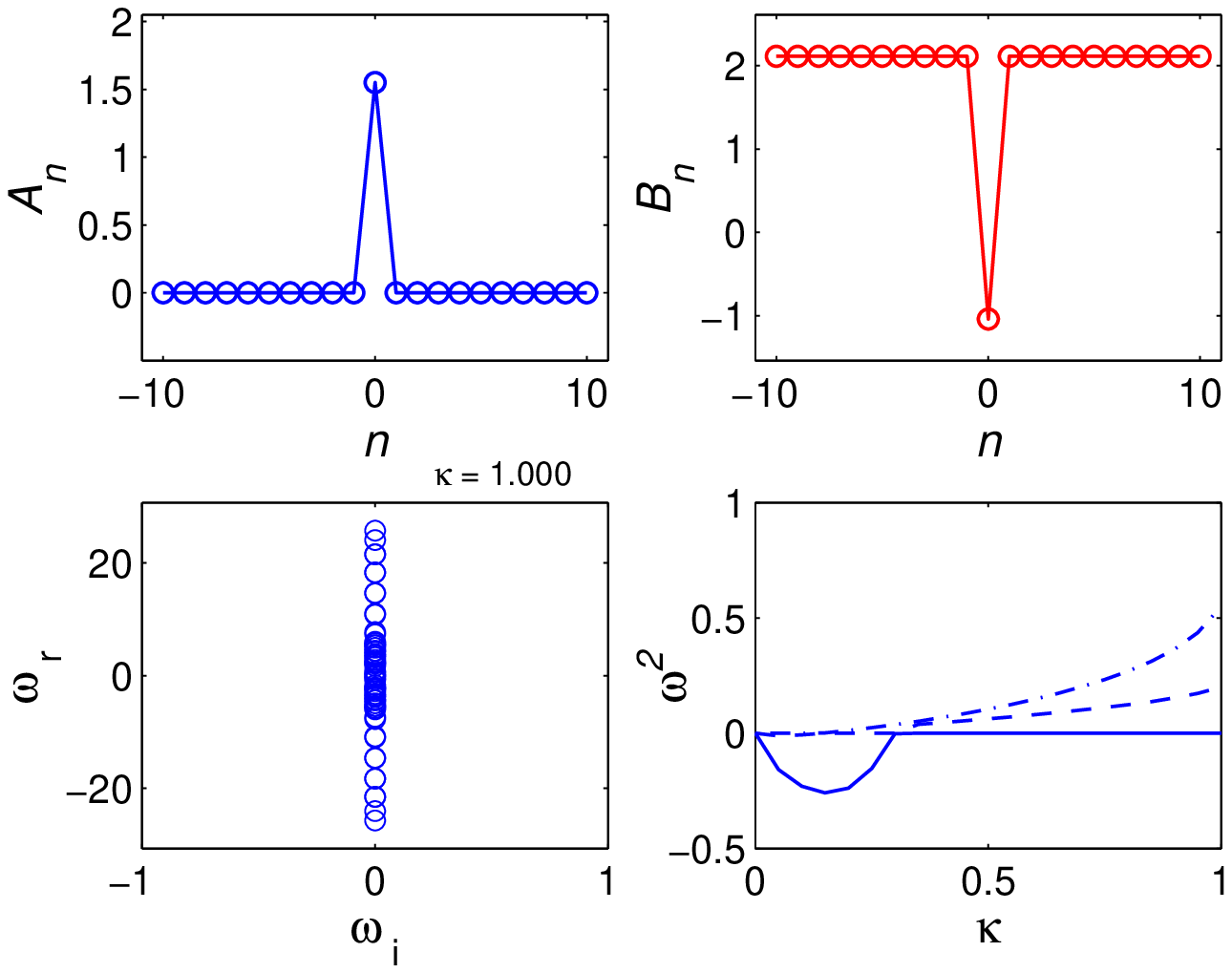}} \centerline{\includegraphics[width=7.5cm,height=3cm]{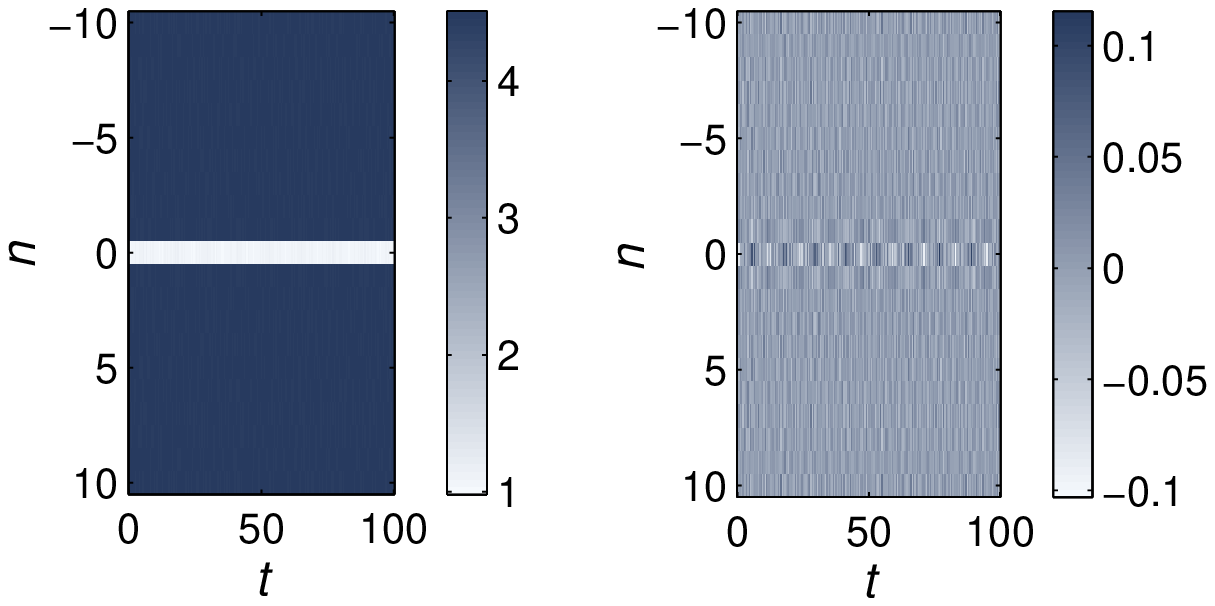}}
\protect\caption{(Color online) First row panels show the amplitude profile and second
row left panel shows eigenfrequency spectrum of one site B-D compactons
for case $\kappa=1$, $\gamma_{1}=1,\:\gamma_{2}=-1$, $\gamma_{12}=-0.5$,
$\alpha_{1}=1$, and $\alpha_{2}=1$. Second row right panel shows
the numerical linear stability analysis as function of $\kappa$.
Bottom panels show the space-time evolution (left panel) obtained
from Eq.~(\ref{vdnlse}), of the dark component (square modulus)
of the compacton depicted in top panels, and its deviation from the
exact stationary solution (right panel). Modulation functions in Eq.~(\ref{vdnlse})
are taken as: $\gamma_{i}=1+\frac{\alpha_{i}}{\epsilon}\cos\left(t/\epsilon\right),\;\; i=1,2,$
with $\epsilon=0.01$.}

\label{brightdarkb}
\end{figure}

We also find that three-site B-B solutions with small $\alpha$ values
display higher discrepancy compared to the respective solution with
large $\alpha$ (by increasing $\alpha$, however, the amplitudes
of the compacton also become smaller). These discrepancies may be
ascribed to the management functions being not sufficiently strong.
It is worth to remark here that in the limit $\epsilon\rightarrow0$
(infinitely strong management) exact compactons of the averaged system
should be exact also for Eq.~(\ref{vdnlse}) (limiting cases of very
small $\epsilon$, however, are quite difficult to investigate numerically
due to the very high accuracy required).

\subsection{Bright-Dark compactons}

It is interesting to look for novel type of solutions of the averaged
DNLSE, which are more characteristic of the two-component systems,
and in particular to B-D (or dark-bright) onsite compactons. Existence
of such solutions follows from the general stationary Eqs. (\ref{mubg1},\ref{mubg2})
by letting $A_{n_{0}}=a,\: B_{n_{0}}=b$, and $A_{n}=0,\: B_{n}=c$
for $n\ne n_{0}$. In this case one finds that exact solutions exist
if chemical potentials and parameters $a,b,c$ of the B-D compacton
are related by the following equations
\begin{equation}
a^{2}=\xi_{0}/\alpha_{1},\;\;\mu_{1}=-a^{2}\gamma_{1}^{0}-c^{2}\gamma_{12},\;\;\ \mu_{2}=-2\kappa_{2}+b^{2}\gamma_{2}^{0},
\end{equation}
\begin{eqnarray}
cJ_{0}((b^{2}-c^{2})\alpha_{2})-b(1+2bc\alpha_{2}J_{1}((b^{2}-c^{2})\alpha_{2}))=0,\nonumber \\
2b\kappa_{2}J_{0}((b^{2}-c^{2})\alpha_{2})+c\left[(c^{2}-b^{2})\gamma_{2}^{0}-2\kappa_{2}+\right.\nonumber \\
\left.\frac{\xi_{0}}{\alpha_{1}}\gamma_{12}+4bc\kappa_{2}\alpha_{2}J_{1}((b^{2}-c^{2})\alpha_{2})\right]=0,\;\;\;\label{bd-eq}
\end{eqnarray}
with $\xi_{0}$ a zero of $J_{0}$. The last equations can be solved
numerically to determine the amplitude $b$ and the background $c$
of the dark component.
\begin{figure}
\centerline{\includegraphics[scale=0.5]{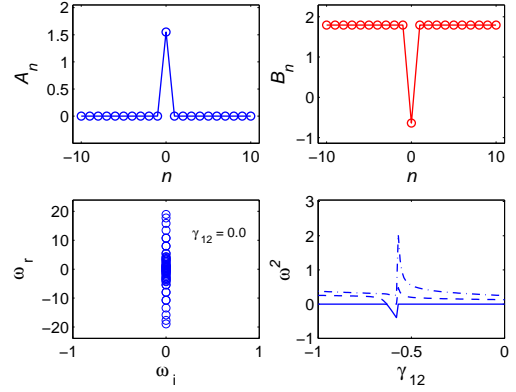}} \protect\caption{(Color online) Same as first two rows panels of Fig.~\ref{brightdarkb},
except for one-site B-D compacton in top panels computed for $\gamma_{12}=0$
and stability properties in the bottom right panel computed as a function
of $\gamma_{12}$. Other parameters are fixed as in Fig.~\ref{brightdarkb}. }

\label{BDg12}
\end{figure}

Typical one site B-D compactons are depicted in Figs.~\ref{brightdarkb},\ref{BDg12}
(see top panels) together with their stability property (see left
middle panels of these Fig.s). Stability properties versus $\kappa$
and $\gamma_{12}$ parameters are displayed in the second row right
panels of Figs.~\ref{brightdarkb},\ref{BDg12}, respectively, from
which we see that in spite of the presence of some instability region,
the range of stability is still quite large. Numerical time evolution
obtained from Eq. (\ref{vdnlse}) and deviation from the exact stationary
dynamics, are investigated for a B-D compacton taken in the left and
right bottom panels of Fig.~\ref{brightdarkb}. The figure refers
only to the dark component, being the bright component in very good
agreement with the stationary solution. Notice from the right bottom
panel the presence of fine periodical ripples on the background and
small oscillations of the deep amplitude in the middle. These deviations,
however, are relatively small compared to dark amplitude and background
of the exact solution, remaining bounded on a long time scale.

\subsection{Dark-Dark Compactons}

Single site dark-dark (D-D) compactons can be searched of the form
$A_{n_{0}}=b_{1},B_{n_{0}}=b_{2}$ and $A_{n}=a_{1},B_{n}=a_{2}$
for $n\neq n_{0}$. The equations for chemical potential and amplitudes
in this case are:
\begin{eqnarray}
\mu_{u} & = & -\gamma_{1}a_{1}^{2}-\gamma_{12}a_{2}^{2}-2\kappa_{1}\nonumber \\
\mu_{v} & = & -\gamma_{2}a_{2}^{2}-\gamma_{12}a_{1}^{2}-2\kappa_{2}
\end{eqnarray}
\begin{eqnarray*}
2\kappa_{1}a_{1}\left(2\alpha_{1}b_{1}J_{1}\left(\xi_{1}\right)+J_{0}\left(\xi_{1}\right)/b_{1}\right)-\gamma_{1}\left(a_{1}^{2}-b_{1}^{2}\right)\\
-\gamma_{12}\left(a_{2}^{2}-b_{2}^{2}\right)-2\kappa_{1}=0\\
2\kappa_{2}a_{2}\left(2\alpha_{2}b_{2}J_{1}\left(\xi_{2}\right)+J_{0}\left(\xi_{2}\right)/b_{2}\right)-\gamma_{2}\left(a_{2}^{2}-b_{2}^{2}\right)\\
-\gamma_{12}\left(a_{1}^{2}-b_{1}^{2}\right)-2\kappa_{2}=0\\
2\alpha_{1}a_{1}b_{1}J_{1}\left(\xi_{1}\right)-b_{1}J_{0}\left(\xi_{1}\right)/a_{1}-1=0\\
2\alpha_{2}a_{2}b_{2}J_{1}\left(\xi_{2}\right)-b_{2}J_{0}\left(\xi_{2}\right)/a_{2}-1=0
\end{eqnarray*}
with $\xi_{i}=\alpha_{i}\left(a_{i}^{2}-b_{i}^{2}\right),i=1,2$.
\begin{figure}
\centerline{\includegraphics[scale=0.5]{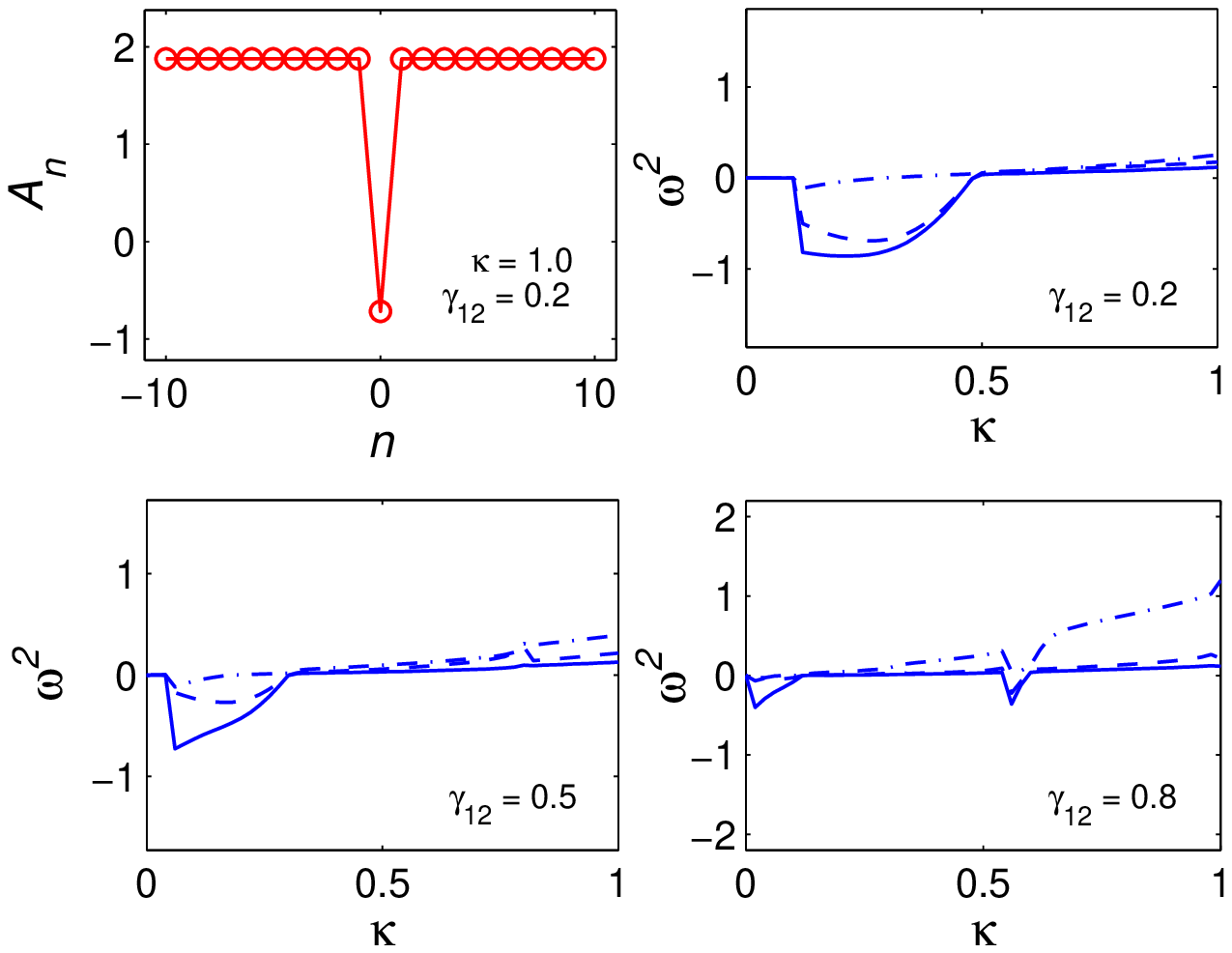}} \centerline{\includegraphics[scale=0.5]{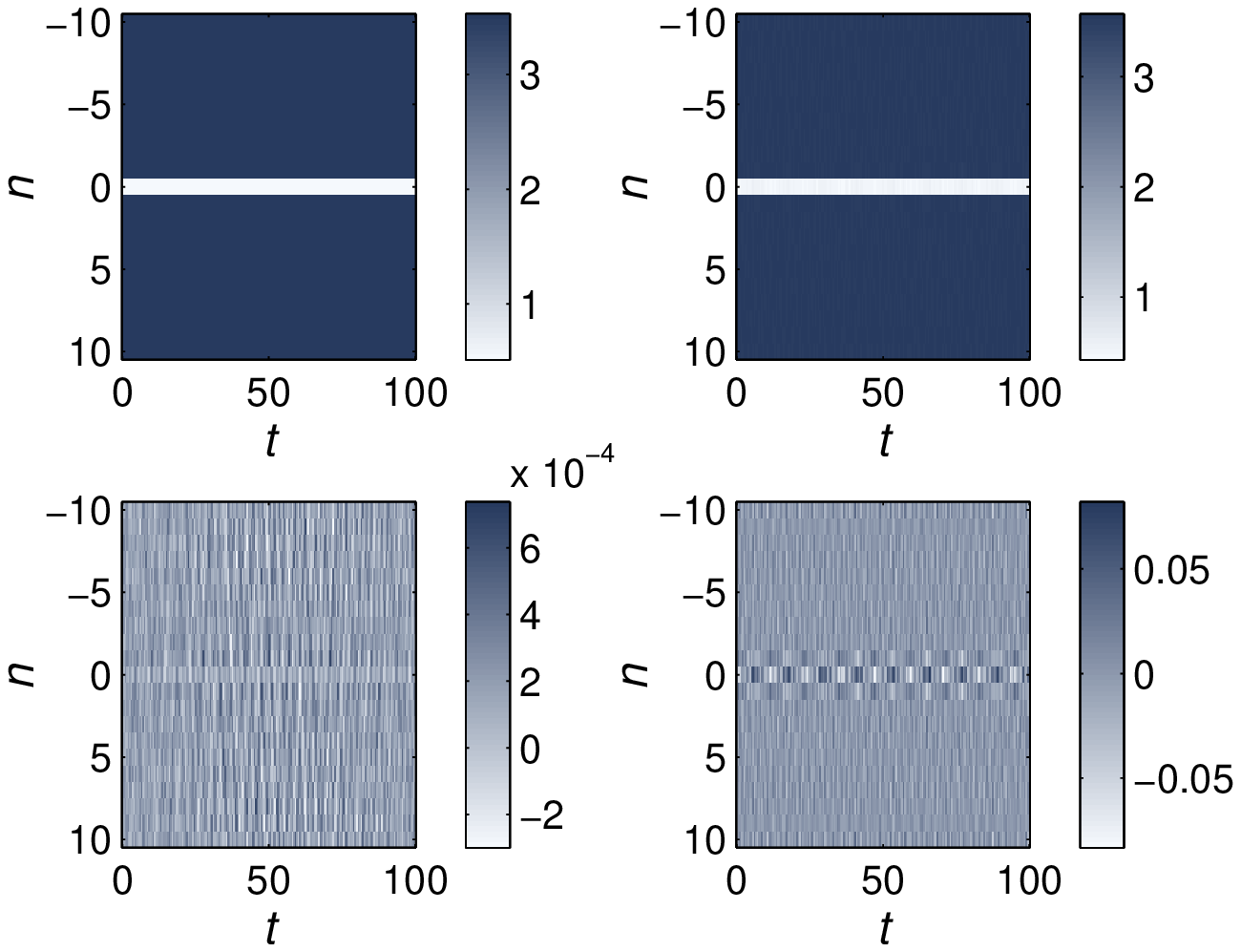}}
\protect\caption{(Color online) First row left panel shows the amplitude profile (identical
for the two components) of a single site D-D compacton for parameter
values $\kappa=1$, $\gamma_{1}=\gamma_{2}=-1$, $\gamma_{12}=0.2$,
$\alpha_{1}=1$, and $\alpha_{2}=1$. In the right panel the numerical
linear stability is reported as as function of $\kappa$. Second row
panels show linear stability analysis for cases $\gamma_{12}=0.5$
(left) and $\gamma_{12}=0.8$ (right). Third row panels show the space-time
dynamics (only one component shown) as obtained from Eqs. (\ref{eq.av1},\ref{eq.av2})
(left) and Eq. (\ref{vdnlse}) (right), for a D-D compacton with parameters
fixed as in top panel but with $\gamma_{12}=0.5$. Bottom panels show
the deviation of the corresponding above dynamics from the exact solution.}
\label{darkdarkstab}
\end{figure}

The first row panels of Fig.~\ref{darkdarkstab} show the amplitude
profile for $\gamma_{12}=0.2$ and respective numerical linear stability
as function of $\kappa$. In the second row panels we report stability
properties for two increasing values of the inter-species interaction:
$\gamma_{12}=0.5,0.8$ (left and right panel, respectively). As one
can see, the $\kappa$ region of stability becomes wider as $\gamma_{12}$
is increased, this indicating a tendency of the coupling to stabilize
dark-dark pairs. Time evolutions with respect to the averaged and
original systems and their deviations from the exact solution are
shown in the bottom two rows left and right panels, respectively.
In general we find that D-D system is more sensitive to disturbance
compared to all previous cases and for higher value of $\kappa$ and
$\gamma_{12}$ (such as $\kappa=1$ and $\gamma_{12}=0.8,$ for example)
much smaller $\epsilon$ and numerical time step are needed to numerically
simulate stable dynamics on a long time scale (exact initial solutions
can be perturbed only by very small noise of the order $10^{-6}$).
A possible reason for this weak stability is discussed in the next
section.

\section{Single component dark compacton}

From the above analysis it appears evident that stability properties
become more critical for compactons that involve dark components.
This is true for B-D and even more for D-D pairs. This fact may result
from a possible instability of the single component dark compacton.
We remark that for single component BEC, compactons were investigated
only for the bright case~\cite{AKS}, so the question of whether
stable single component dark compactons can exist, is open. In this
section we provide the answer to the question by studying the uncoupling
limit $\gamma_{12}\rightarrow0$ of the D-D compacton discussed before.

\begin{figure}
\centerline{ \includegraphics[scale=0.5]{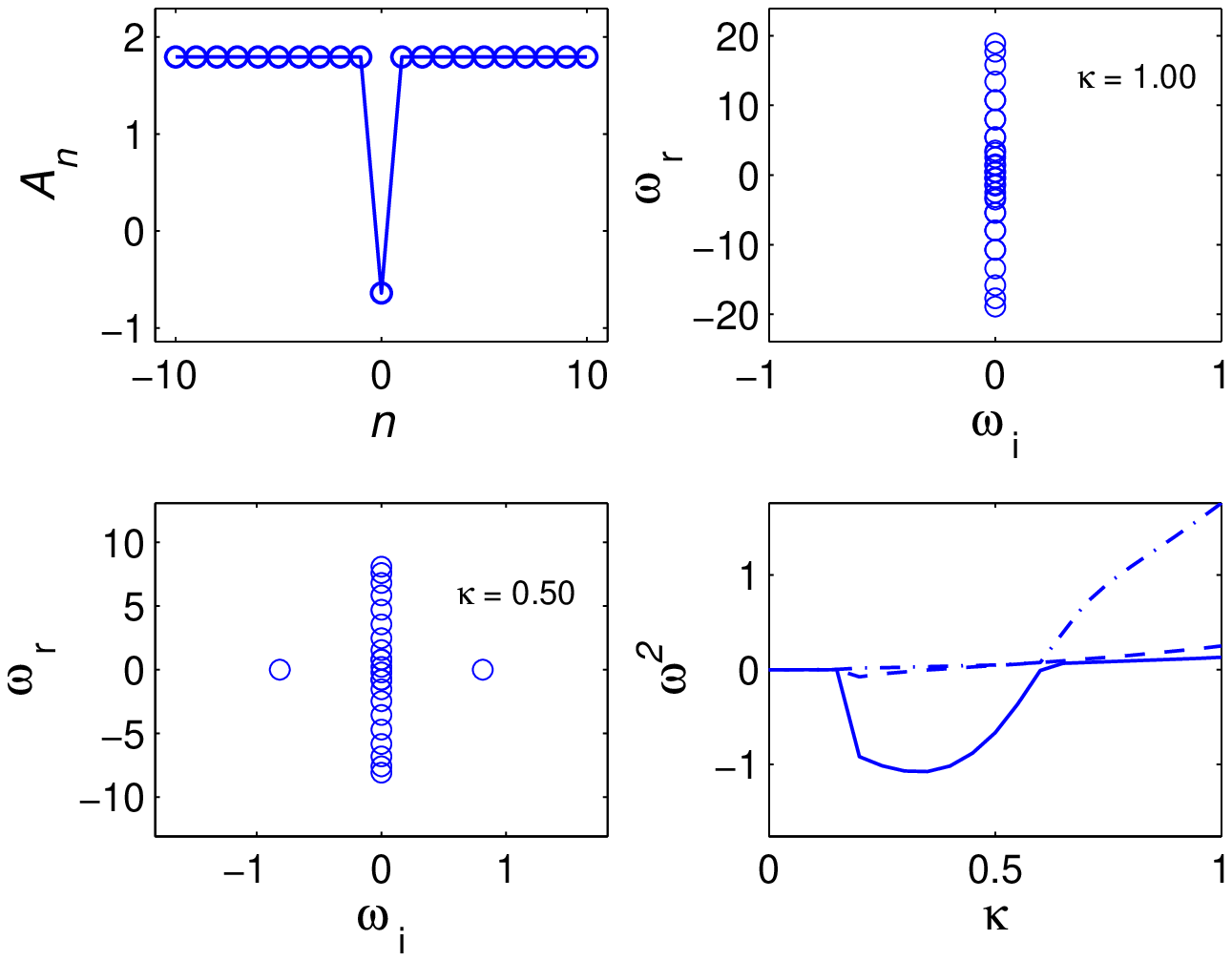}} \centerline{\includegraphics[scale=0.5]{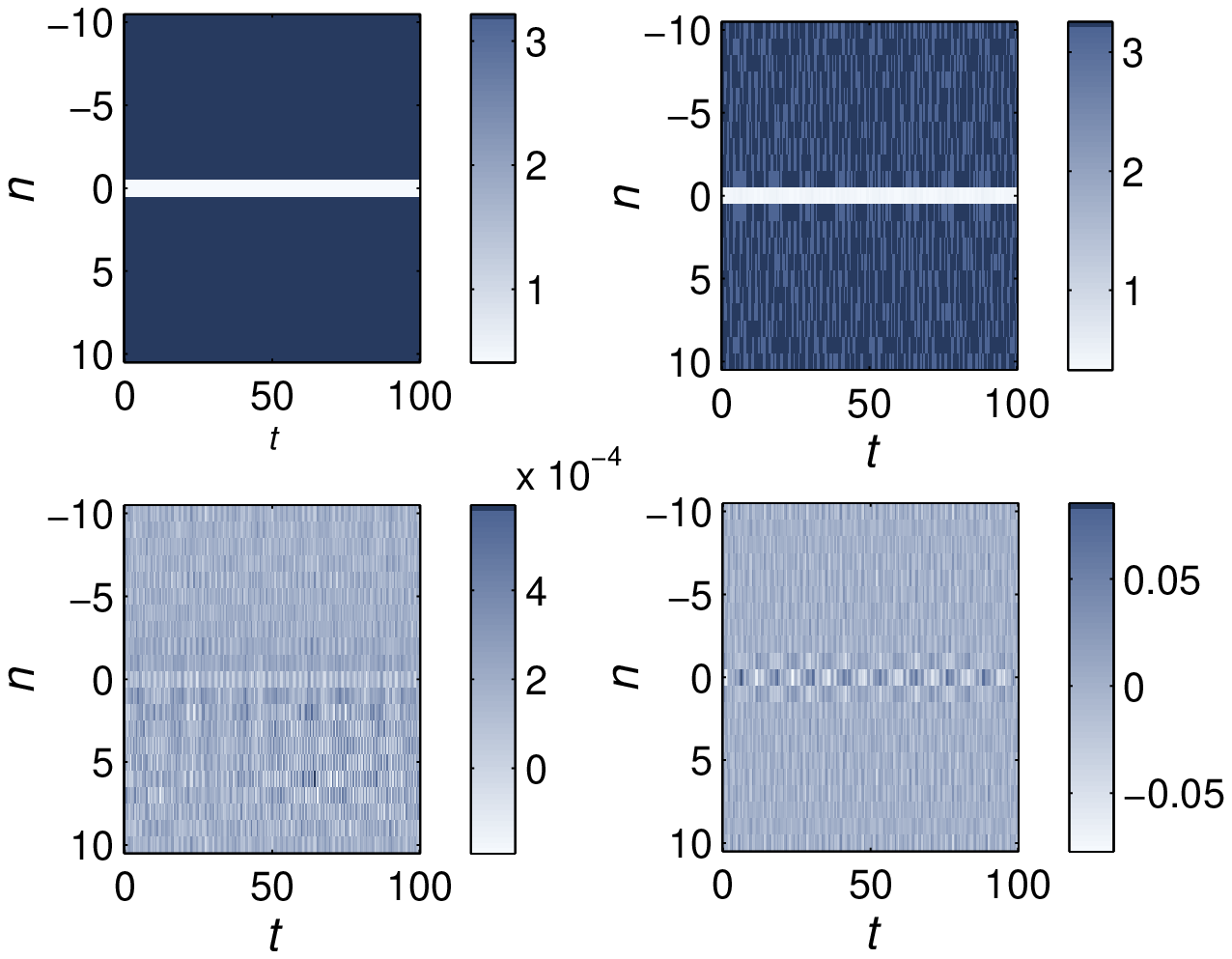}}
\protect\caption{(Color online) First row panels show the amplitude profile (left)
and eigenfrequency spectrum (right) of uncoupled one site dark compacton
for case $\kappa=1$, $\gamma_{0}=1$ and $\alpha=1$. Second row
panels show eigenfrequency spectrum for case $\kappa=0.5$ (left)
and the numerical linear stability analysis as function of $\kappa$
(right). Last two row panels. Space-time evolution of one site dark
compacton solution for parameter case as in upper panels. Third row
panels show the square modulus of the solution, the left side from
the averaged system Eq.(\ref{eq.av1},\ref{eq.av2}) and the right
side from Eq.(\ref{vdnlse}), while last row panels show their respective
deviations from the exact solution.}

\label{darkcase}
\end{figure}

In this respect the averaged equations in (\ref{eq.av1},\ref{eq.av2})
for $\gamma_{12}=0$ with $\kappa_{i}\equiv\kappa,\:\alpha_{i}\equiv\alpha,\:\theta_{i,\pm}\equiv\theta_{\pm}$
reduce to the same equations considered in~\cite{AKS} for the single
field variable $U_{n}=V_{n}\equiv u_{n}$. Considering $u_{n}=A_{n}\exp(-i\mu t)$
then the equation becomes
\begin{eqnarray}
\mu A_{n}=-2\kappa\alpha A_{n}^{2}\left[A_{n+1}J_{1}(\alpha\theta^{+})+A_{n-1}J_{1}(\alpha\theta^{-})\right]\nonumber \\
-\kappa\left[A_{n+1}J_{0}(\alpha\theta^{+})+A_{n-1}J_{0}(\alpha\theta^{-})\right]+\gamma_{0}A_{n}^{3}\label{scstat}
\end{eqnarray}
with $\theta^{\pm}=A_{n\pm1}^{2}-A_{n}^{2}$. Dark modes are possible
only for repulsive interactions so that in the above equation $-\gamma_{1}^{(0)}\equiv\gamma_{0}>0$.
A dark single site compacton located at the $n=n_{0}$ site can be
determined by setting $A_{n}=b$ if $n=n_{0}$ and $A=a$ for $n\neq n_{0}$
to yield
\begin{eqnarray}
 &  & 2a\kappa J_{0}(\xi)+b[(a^{2}-b^{2})\gamma_{0}-2\kappa+4ab\kappa\alpha J_{1}(\xi)]=0\nonumber \\
 &  & bJ_{0}(\xi)-a(1+2ab\alpha J_{1}(\xi))=0\label{mubg}
\end{eqnarray}
where $\xi=\alpha\left(a^{2}-b^{2}\right)$. It follows that to satisfy
the equation at the nonvanishing sites, one must have
\begin{equation}
\mu=a^{2}\gamma_{0}-2\kappa.\label{mu}
\end{equation}
then Eq.(\ref{mubg}) can be numerically solved to obtain $\{a,b\}$.

Numerical linear stability analysis for parameters values $\gamma_{0}=1$
and $\alpha=1$ (see second row, right panel of Fig.~\ref{darkcase})
shows that the solutions are unstable until $\kappa>0.65$. A comparisons
with the first two rows of Fig.~\ref{darkdarkstab} indicates that
the stability regime can be improved by inter-species coupling in
the two component DNLS system. Nevertheless, when $\kappa>0.65$,
stable dark compactons exist as shown in the third row of Fig.~\ref{darkcase}
for case $\kappa=1$.

From this it appears evident that, in contrast with the bright compactons
case, the strong nonlinear management does not provide (at least for
the simple real amplitudes ansatz assumed) stable single component
dark compacton solutions for wide range of parameters. It is interesting
that this situation is slightly improved when the inter-species interaction
is switched on, as discussed before.

\section{Generation of binary BEC compactons and experimental setting}

It is also interesting to discuss physical conditions for which compactons
could be experimentally observed in ultracold BEC mixtures. From a
first sight one could think that, in order to keep the zero tunneling
condition satisfied at the edges one needs a very precise control
of the number of atoms inside the compacton, a fact that could be
difficult to arrange in a real experiments. On the other hand, if
the excitation is very stable, it should appear also in the presence
of generic fluctuations that are unavoidable in any real experiment.
In this respect is of interest to address the problem of compacton
generation from generic initial excitations, this providing evidence
of their robust emergence even under unfavorable conditions. To this
end, let us concentrate on the most stable compact excitations, which,
as we have seen, are the ones of B-B type. Without loss of generality,
we take initial excitations of Gaussian-type for both components $u_{n}\left(0\right)=A\exp(-\eta\, n^{2})/\alpha_{1},\; v_{n}\left(0\right)=A\exp(-\eta\, n^{2})/\alpha_{2}$,
with $A,\eta$ fixed as $A=1.55,\eta=0.2$ in the following numerical
simulations (similar results can be obtained with other types of initial
conditions).

\begin{figure}
\centerline{\includegraphics[scale=0.5]{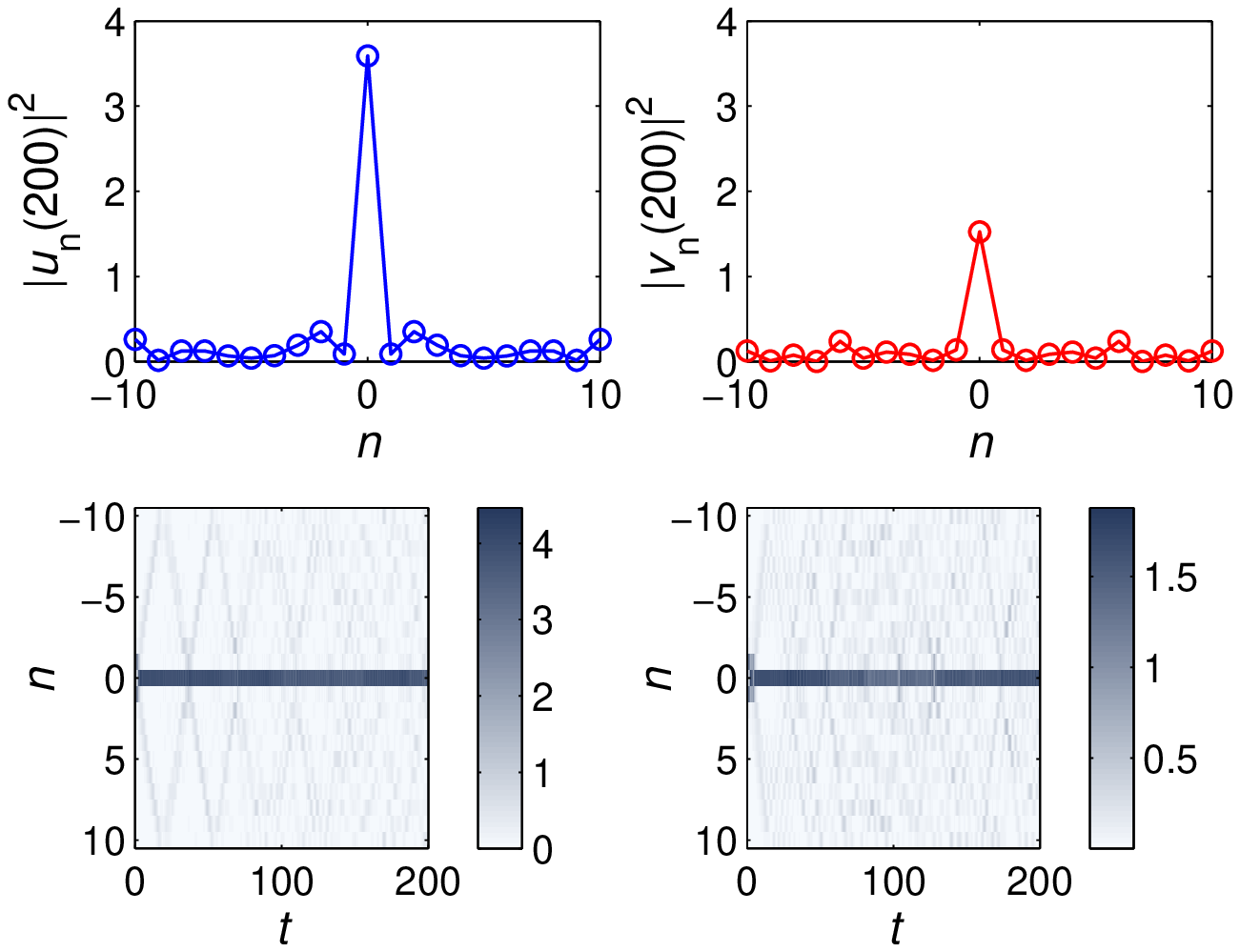}} \centerline{\includegraphics[scale=0.5]{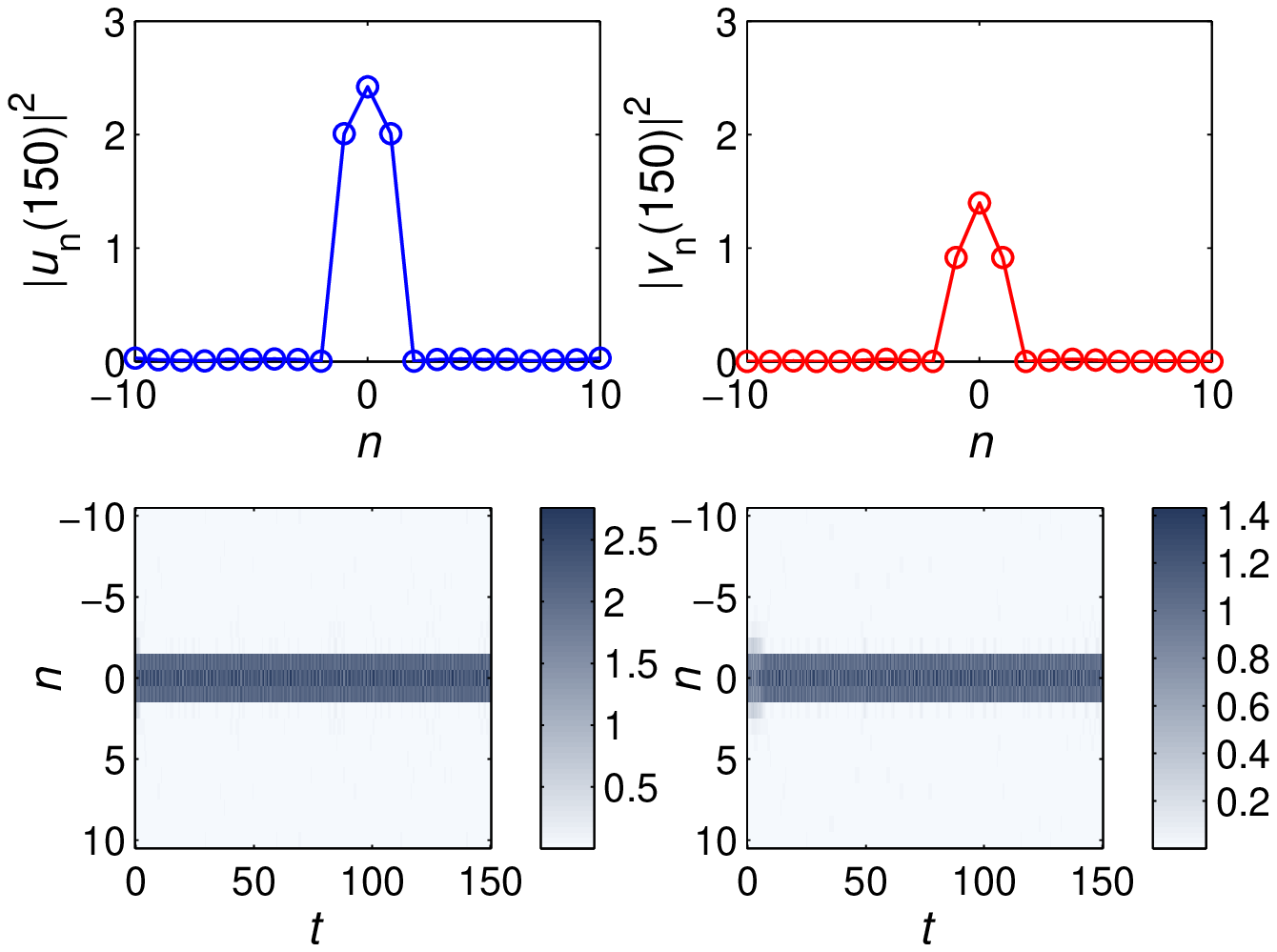}}
\protect\caption{(Color online) Top two row panels. Single site B-B compacton emerging
from the time evolution of an initial Gaussian excitation (described
in the text) of Eq.~(\ref{vdnlse}) with nonlinear management function
taken as in Fig.~\ref{brightdarkb} but with $\epsilon=0.1$. Parameter
are fixed as $\kappa=0.5$, $\alpha_{1}=1$, $\alpha_{2}=2$, $\gamma_{1}=\gamma_{2}=1,\gamma_{12}=0.5$.
Top panels show the density profiles of the first (left) and second
(right) component at time $t=200$ while second row panels show the
corresponding time evolutions. Bottom two row panels. Same as in top
two row panels but for $\epsilon=0.01$, and $\kappa=1$ and for a
three site B-B compacton displayed at time $t=150$ . Other parameters
are as in top two row panels.}
\label{bbg12}
\end{figure}

In Fig.~\ref{bbg12} we show the generation of a single site (see
top two row panels) and a three sites (see bottom two row panels)
B-B compacton from a Gaussian initial condition (same for both cases).
The first and third row panels refer the density profiles (at time
$t=200$ and $t=150$, respectively) while second and forth row panels
refer to time evolutions of the two components. From top two row panels
we see that after expelling some excess matter, a single site compacton
on site $n=0$ (plus some background noise), emerge. The background
noise is unavoidable because the number of atoms (squared amplitudes)
in the compacton for the chosen parameter is fixed by the first zero
of $J_{0}$ and is smaller than the number of atoms in the initial
Gaussian. In spite of this, the densities of the two components in
the neighboring sites, $n=\pm1$, keep very small (although not exactly
zero) during their time evolution, irrespectively from the excess
radiation. For the considered case, $\epsilon$ was $0.1$ and therefore
the management was not very strong. One can expect that by further
decreasing $\epsilon$ one can better and better approximate a true
compacton solution with vanishing densities at $n=\pm1$ (this becoming exactly true in the limit $\varepsilon\rightarrow0$).
The effect of a stronger management can be see from the bottom two
row panels of Fig.~\ref{bbg12}. In this case epsilon is one order
of magnitude smaller ($\epsilon=0.01$), $\kappa=1$ and the mismatch
between numbers of atoms in the initial and final states is quite small.
As a result, we see that a three site compacton, practically indistinguishable
from an exact solution based on the first zero of $J_{0}$, is formed.
This clearly demonstrates that, with proper parameter design and proper
management conditions, compactons can be very robust excitations that
can emerge spontaneously from generic initial conditions.

\begin{figure}
\centerline{\includegraphics[scale=0.5]{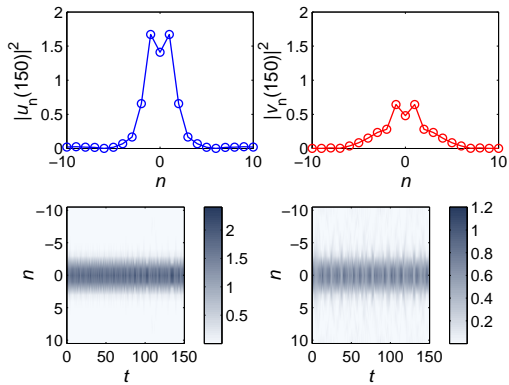}} \protect\caption{(Color online) Same as in bottom two row panels of Fig.~\ref{bbg12}
but for $\gamma_{12}=0.2$.}
\label{ilm-tail}
\end{figure}

Further results shown in Fig.~\ref{ilm-tail} address the problem
of the influence of the inter-species interaction $\gamma_{12}$
on the compacton formation . Here same parameters and same Gaussian
initial excitations as in bottom two row panels  in Fig.~\ref{bbg12} are used,
but with lower value of the inter-species interaction, $\gamma_{12}=0.2$.
 In this case we see that instead of a three site compacton,
an intrinsic localized mode with exponential tail is formed, in full
agreement with the fact that by lowering the inter-species interaction
the compacton stability  may be lost.

Let us now discuss possible experimental setting for the observation
of compacton modes. In this respect we remark that the existence bound
pairs of B-D solitons (localized excitation with tails) has been experimentally
demonstrated in~\cite{Shrestha} for the continuous case. A possible
experimental setting to observe the corresponding compacton modes
could be a BEC mixture of $^{41}$K and $^{87}$Rb atoms loaded in
a deep optical lattice considered in the experiment~\cite{Catani}
and subjected to inter-species scattering length modulations. An equivalent
setting could be implemented also in nonlinear optics with an array
of optical wave-guides with varying Kerr nonlinearity along the propagation~\cite{ATMK,Assanto}.
For the case of binary BEC mixtures, the time modulations of the scattering
lengths can be easily implemented by means of the FR technique by
varying the external magnetic field, $B$, near a resonant value:
\begin{equation}
a_{ij}(t)=a_{ij,br}(1-\frac{\Delta}{B_{ij,0}-B(t)}).
\end{equation}
Here $a_{ij,br}$ denotes the background value of the respective scattering
lengths and $B_{ij,0}$ the resonant value of the magnetic field.
In the case of $^{87}$Rb atoms, for example, there are several narrow
FR which could be used, the broadest one lying at $B_{0}=1007$G.
The tight-binding limit, appropriate for a vector DNLSE description,
could be reached by considering optical lattices of amplitude $V_{0}>10E_{R}$,
where $E_{R}=\hbar^{2}k^{2}/2m$ is the recoil energy. Thus, by changing
periodically and rapidly in time the magnetic field around a FR of
the intra-species scattering lengths 
it should be possible to observe two-component matter wave compactons
to emerge from generic initial conditions in real experiments.

\section{Conclusions}

In conclusion, we have investigated the existence and stability of
binary mixtures matter waves in arrays of BEC subjected to time dependent
periodic variations of the scattering length, by means of an averaged
two- component DNLSE. In addition of B-B compactons we showed that
B-D and D-D compactons are also possible. The stability of
these modes has been investigated both by linear spectral analysis
and by direct numerical integrations. We found that the single site
and the two-site (out-of-phase) B-B compactons are always very stable
in the whole parameter range, while for the other modes there exist
thresholds in the tunneling constant rate below which they cannot
exist as stable excitations. The stability resulted in general to
be more critical for pairs involving one (or both) component of dark
type. In all cases, however, the predictions of the averaged system
were found in good agreement (in some cases; excellent) with the results
of the numerical simulations.

\section*{Acknowledgements}

M. S. acknowledges partial support from the Ministero dell'Istruzione,
dell'Universitá e della Ricerca (MIUR) through a PRIN (Programmi di
Ricerca Scientifica di Rilevante Interesse Nazionale) 2010-2011 initiative.

\end{document}